\documentclass[aps,prd,10pt,onecolumn,nofootinbib,showpacs,showkeys,notitlepage]{revtex4-1}
\usepackage{amsmath,graphicx,bm}
\usepackage{hyperref}

\renewcommand{\d}{\partial}

\newcommand{\nn}{\nonumber\\}

\newcommand{\rh}{\varrho}

\newcommand{\exv}[1]{\left\langle{#1}\right\rangle}

\newcommand{\p}{{\bf p}}

\renewcommand{\Im}{\,\textrm{Im}\,}

\newcommand{\pint}[2]{{\int\!\frac{d^{#1}#2}{(2\pi)^#1}\,}}
\newcommand{\pintz}[1]{{\int\!\frac{d #1}{2\pi}\,}}

\renewcommand{\c}[1]{{\cal{#1}}}

\newlength{\szovszel}\newlength{\szovmag}

\newcommand\lsim{\mathrel{\rlap{\lower4pt\hbox{\hskip1pt$\sim$}} \raise1pt\hbox{$<$}}}                % less than or approx. symbol
\newcommand\gsim{\mathrel{\rlap{\lower4pt\hbox{\hskip1pt$\sim$}} \raise1pt\hbox{$>$}}}                % greater than or approx. symbol

\newcommand{\K}{{\cal K}}

\begin{document}

%\title{QCD thermodynamics: strongly correlated quasiparticles}
%\title{QCD pressure: gradually melting hadrons until 3$T_c$}
\title{QCD over $T_c$: hadrons, partons and continuum}

\author{T.S. Bir\'o} 
%\email{tsbiro@wigner.hu}
\affiliation{MTA Wigner Research Centre for Physics, H-1525 Budapest, P.O.Box 49, Hungary}

\author{A. Jakov\'ac} 
%\email{jakovac@phy.bme.hu}
\affiliation{Institute of Physics, E\"otv\"os University, H-1117 Budapest, Hungary}
%\affiliation{Institute of Physics, Eotvos University, H-1117 Budapest, Hungary}

\date{\today}

\begin{abstract}
  In this paper we provide a physical picture for the QCD phase
  transition in terms of qualitative changes in the spectral functions. 
  Our approach takes into account the crossover nature of this transition
  and counts for the observed strong correlation seen in higher order
  susceptibilities. We demonstrate that the hadron resonance gas,
  which alone describes the thermodynamics at temperatures $T<T_c$,
  will appreciably contribute to the total pressure until $T\lsim
  3T_c$. In this intermediate regime the QCD matter consists of
  strongly correlated excitations, interpretable as either hadrons or partons. 
  As hadronic spectral peaks gradually vanish, the partonic excitations
  start to form a stand-alone quasiparticle gas. 
  The conventional picture of a quark gluon plasma emerges only at $T\gsim 3T_c$.
\end{abstract}

\maketitle

\section{Introduction}
\label{sec:intro}

The strongly interacting elementary matter is described by Quantum
Chromodynamics (QCD). As it was proven by Monte Carlo (MC) measurements, 
it changes from hadronic matter to quark-gluon-plasma (QGP) phase by a continuous 
transition. One distinguishes a (would-be) critical temperature
$T_c$ located at the peak of some susceptibilities: e.g. $T_c=150$ MeV
for chiral susceptibility~\cite{Aoki:2006we}, for other observables
one finds somewhat different values~\cite{Borsanyi:2010bp}. It is
remarkable, that from strange quark susceptibility one already obtains a
significantly (about 15\%) higher crossover temperature~\cite{Bellwied:2013cta}.

The thermodynamics at high temperatures can be reproduced using the
fundamental QCD degrees of freedom, the quarks and gluons.  For
obtaining the pressure we need to perform resummation, recently a
three-loop Hard Thermal Loop (HTL) level~\cite{Haque:2014rua}
has been achieved.  In this way one obtains values 
describing the measured Monte Carlo pressure at $T\gsim 2T_c$ satisfactorily.
At low temperatures, up to $T\lsim T_c$, the thermodynamical
description is based on the non-interacting gas of hadron
resonances~\cite{HRG0} (Hadron Resonance Gas, HRG). In this description
a free gas of hadrons is taken, with masses determined according
to the Particle Data Group~\cite{PDG} tables, and the pressure is
calculated by the free pressure formula. This simple picture can
reproduce different measurements in MC 
studies~\cite{Andronic:2003,Karschetal,Huovinen:2009yb,Borsanyi:2010cj,Bazavov:2013dta}
up to $T_c$ surprisingly well.

The $T_c<T<(2\,-\,3)T_c$ regime (crossover regime) is  much less understood.
First of all, since the phase transition is a crossover, we
expect that all thermodynamical quantities change continuously. 
This, combined with the fact that HRG describes
the QCD pressure well at $T_c$, implies that the hadrons
do not disappear from the thermal ensemble at $T>T_c$~\cite{Liao:2005pa}. 
In fact, direct MC measurements support this 
idea~\cite{Datta:2003ww,Umeda:2002vr,Asakawa:2003re, Jakovac:2006sf,Petreczky:2012ct}. 

The success of HRG demonstrates that, at least at low temperatures, the
strongly interacting QCD is dual to a weakly interacting hadron
model. Put another way, the bulk effect of interactions is
realized by the modification of the spectral functions, and the newly
formed quasiparticles are almost interaction-free. Since at $T_c$ the
HRG description seems to work, the hadronic interaction must be weak
for $T>T_c$. This means that the main properties of the crossover regime 
can be described by a weakly or even non-interacting mixture of hadrons, 
quarks and gluons. We must be aware, however, that in the language of the 
original degrees of freedom the interaction is rather strong, and 
therefore we shall expect strong deviations from the most naive small width 
quasiparticle description. 

A first approximation for this spectrum modification is to assume
that the \emph{masses} of the thermodynamical degrees of freedom are
modified~\cite{Liao:2005pa}. At high temperature the hadronic masses,
at low temperature the quark masses are growing rapidly. Since their
thermal weight is diminishing, we do not see hadrons at high and
quarks at low temperature. Unfortunately, this first approximation
fails. The immediate problem is that we do not see this tendency
in the MC spectra~\cite{Maezawa:2013nxa, Iida:2010jz}. Numerical
findings are consistent with a thermal mass satisfying $m_{{\rm therm}}^2=m_0^2+cT^2$, but
nothing particular happens at $T_c$. There are, moreover, indirect
arguments, too. One may measure correlator combinations on the lattice
that are zero (or constant) in an uncorrelated quasiparticle model~\cite{Bazavov:2013dta}. 
The measurement of such correlators~\cite{Bazavov:2013dta, Bellwied:2013cta} 
show a clear distinction from an uncorrelated model. 
Another indirect proof for the insufficiency of considering
quasiparticle mass changes only is that in the
small-width, weakly interacting quasiparticle model the transport
coefficients are large~\cite{transportcoeff}, while in the heavy ion
collisions one observes rather small viscosity to
entropy ratios, $\eta/s$~\cite{Heinz:2013th}.

We conclude that the excitation spectrum is modified more than just by
a simple mass shift. In fact we expect that a realistic
quasiparticle spectrum consists of a quasiparticle peak, characterized
by its mass, width and wave function renormalization, and of the continuum
part. Unfortunately we do not have access to the continuum of
the different quantum channels in QCD, neither from experiments, nor
from lattice studies. Perturbative calculations are of limited use in this
regime. There are, on the other hand, some expectations which must be true in
any models, and also seen in experimentally measured spectra~\cite{Arnaldi:2006jq}. 
Namely while the continuum part must be nonzero for all
four-momenta at finite temperatures, we expect it to be enhanced
above some threshold value. Apart from occasional resonances we do
not anticipate any fast change in the continuum.

Several authors have studied the effect of the spectrum on
the thermodynamics in some 
approximation~\cite{Blaschke:2003ut,Biro:2006zy,Biro:2006sv,Jakovac:2013iua} 
and also on transport coefficients~\cite{Ivanov:1998nv,Ivanov:1999tj,Peshier:2005pp,NoronhaHostler:2012ug}. 
According to general experience quasiparticle peaks yield similar
contribution to the thermodynamics as free particles, but the
presence of the continuum appreciably reduces the pressure.
Thus in order to describe the pressure of QCD we should consider even in a 
first approximation hadrons as well as partons (i.e. quasiparticles
formed by quarks and gluons) with a generic spectral function. In these
spectral functions the temperature variation of the masses must be
moderate~\cite{Maezawa:2013nxa,Iida:2010jz}, in accordance with MC
measurements. The main effect is the change in the width
and height of the quasiparticle peaks and the height of the
continuum as the temperature varies. These are not fully
independent quantities, they are connected by the sum rules.

The main goal we aim at in this paper is finding a
complete description of QCD thermodynamics based on the spectral
representation of the QCD excitations. Having this we ask then physical
questions, like how fast will the hadrons disappear from the $T>T_c$ plasma? 
For the mathematical details we examine typical spectral
functions as inputs and build effective quadratic models which are
compatible with the given spectral functions. 
Using the formulae of~\cite{Jakovac:2012tn,Jakovac:2013iua} we determine the
thermodynamics based on these spectral functions. The model spectral
functions contain several peaks and one continuum, thermal mass and
spatial momentum dependence. We seek for a method to
represent the resulting pressure in some common way. 
This happens by using the notion of the effective number of degrees of freedom, $N_{eff}$,
which describes the ratio of the actual pressure to the ideal pressure.
The decrease of this effective number of degrees of freedom
represents the melting of the given excitation.

The findings of this study will then be applied to describe QCD
pressure. We propose to use a statistical approach for the description of hadrons
and a quantum -- correlated description for the partons. This latter means that
the abundance of the partons depends very strongly on the
available hadronic particles: whenever the system is full with hadrons, the
partons must have short lifetimes.

The result of our analysis can be summarized in some simple
sentences. First of all, due to the melting and correlation effects,
the full QCD pressure can be exhausted by our model, moreover the
partial pressures of hadrons and partons are obtained separately.  We
find the hadrons responsible for the full pressure below $\sim T_c$,
they dominate the pressure for $T\in[T_c,2T_c]$ and vanish as
thermodynamical degrees of freedom around $\sim3T_c$.  For the partons
we obtain the reverse story: they are the only thermodynamical
ingredients for $T\gsim 3T_c$, dominate the pressure for
$T\in[2T_c,3T_c]$, and vanish from the ensemble at $T\sim T_c$.  We
realize furthermore that in the intermediate temperature interval {\em
  the excitations are not fully particle-like}, their pressure
contribution is considerably below the free gas pressure. This
conclusion is supported by recent correlation measurements in lattice
QCD~\cite{Bazavov:2013dta, Bellwied:2013cta}.

This paper is organized as follows: in Section~\ref{sec:genspect} we
establish how a typical spectral function of QCD excitations should be parametrized.
Then one-by-one we examine the dependence of the effective
number of degrees of freedom on the relevant parameters of the
spectrum: on the quasiparticle width and height in subsection~\ref{sec:QPwidth}, 
on the spatial momentum dependence in subsection~\ref{sec:momdep} 
and on the thermal mass dependence in subsection~\ref{sec:thermmass}. 
In Section~\ref{sec:QCDthermo} we use these
findings for building an effective statistical model of QCD, containing melting
hadrons and correlated partons as constituents. We compute the
pressure and analyze the result in~\ref{sec:discussion}. We close this
paper with our Conclusions (Section~\ref{sec:conc}).

%%%%%%%%%%%%%%%% SECTION II %%%%%%%%%%%%%%%%%%

\section{Realistic description of excitations in QCD}
\label{sec:genspect}

The goal in the present Section is to characterize a generic spectral
function by some physically relevant parameters, and to compute the corresponding pressure.
This technique is closely related to resummation techniques, where
one also parametrically modifies the spectrum of excitations. From this point
of view the HRG description itself can be considered as a
nonperturbative resummation Ansatz, where we use our knowledge of the
low energy QCD spectrum, and single out the most prominent property,
the masses. Being an approximation, the success is not
guaranteed, but it happens to work for $T<T_c$ quite well. Then we can
tell that, by experience, the most relevant degrees of freedom of QCD
are the hadrons. At high temperature, $T>(2-3) T_c$ the QCD degrees of
freedom, after resummation, can also describe the measured QCD
pressure. 

The temperature interval $T\in[T_c,(2-3)T_c]$ seems to be hardly
accessible either for the QCD resummation method and for the most naive
nonperturbative hadronic excitation ``resummation''. QCD resummations
are unreliable due to the large gauge coupling, while HRG overshoots
the pressure above $T_c$ (in fact it diverges to infinity at the
Hagedorn temperature). Moreover, it is not known how the different
resummation methods could join to each other smoothly, as dictated by
the observed crossover phase transition. And finally MC measurements
suggest considerable correlation between the particle species,
suggesting that the free particle description is rather far from the
truth.

But before we label this region as strongly interacting and
perturbatively not accessible, we should first try to improve the
Ansatz for the nonperturbative hadronic ``resummation'' model beyond
the most naive choice of HRG. We maintain the property that it
describes non-interacting excitations, but we allow for the most
generic hadron, as well as quark and gluon spectra. Unfortunately we do
not have access for a real hadronic spectral function, neither from
experiments, nor from numerical simulations (although there are some
results there). On the other hand we know that near $T_c$ in the HRG
model one has to take into account the contribution of ${\cal O}(2000)$ 
hadronic resonances. Therefore what is needed is not the
spectrum of each single hadron, which may differ from
each other, but only a statistical description. We have to know only
the ``typical'' spectrum. For that we already have a good guess, since in
all model calculations a typical spectrum consists of one or more
quasiparticle peaks and a continuum. The height of the continuum
depends on the number of available scattering channels as well as on
the coupling strength to them, while the complete spectral function is
subject to generic sum rules. The masses of the quasiparticles at
different quantum channels also should be chosen statistically, the
Hagedorn-spectrum is a good candidate for
that~\cite{Jakovac:2013iua,Hagedorn:1965st,Broniowski:2004yh}.

We still need a method to calculate thermodynamics once the
spectrum is given. This procedure has already been worked out in~\cite{Jakovac:2012tn}. 
The key result is that if we have a spectral
function $\rh$ then the pressure of the system is given as
\begin{equation}
  \label{eq:pressure}
  P = -\alpha T \pint4p \Theta(p_0) \frac{D\K_\rh(p)}{p_0}\, 
  \ln\left(1- \alpha e^{-\beta p_0}\right) \rh(p),
\end{equation}
where $\alpha=\pm1$ for bosonic/fermionic modes. The function
$D\K_\rh(p)$, besides its momentum dependence, also depends on the
spectral function:
\begin{equation}
   \K_\rh(p)  = \left({\cal P}\pintz\omega
     \frac{\rh(\omega,\p)}{p_0-\omega}\right)^{-1},\qquad  D\K_\rh(p)
   = p_0 \frac{\d\K_\rh}{\d p^0}\, -\,\K_\rh,
\end{equation}
where ${\cal P}$ denotes the principal value integral. Physically $\K_\rh$
represents the kernel of the effective model reproducing the spectral
function $\rh$.

The consistency of this description is proven in~\cite{Jakovac:2012tn}. 
Here we mention that this theory is a consistent
field theory, it is unitary, causal, energy and momentum conserving, and (if
needed) Lorentz-invariant. Whenever the spectral function consists of $N$
separate Dirac-delta peaks (with possibly unequal heights), then
the pressure formula reproduces the well known bosonic/fermionic free
gas expressions with $N$ constituents. One realizes
that the pressure is independent of the overall normalization (sum
rule). 

These formulae use the spectral function as an input. The spectral
function is a hardly accessible quantity. In general, at finite
temperature we expect a dependence of
\begin{equation}
  \label{eq:rho}
  \rh(p) = \rh_Q(p_0,|\p|; T,\mu).
\end{equation}
The index $Q$ symbolizes the quantum channel: naturally the spectrum depends
on the quantum numbers. At finite temperature Lorentz invariance is
broken, and so we should have a separate dependence on $p_0$ and
$p=|\p|$. The spectral function also depends on the environmental
variables, like $T$ and $\mu$.

\begin{figure}[htbp]
  \centering
  \includegraphics[height=5cm]{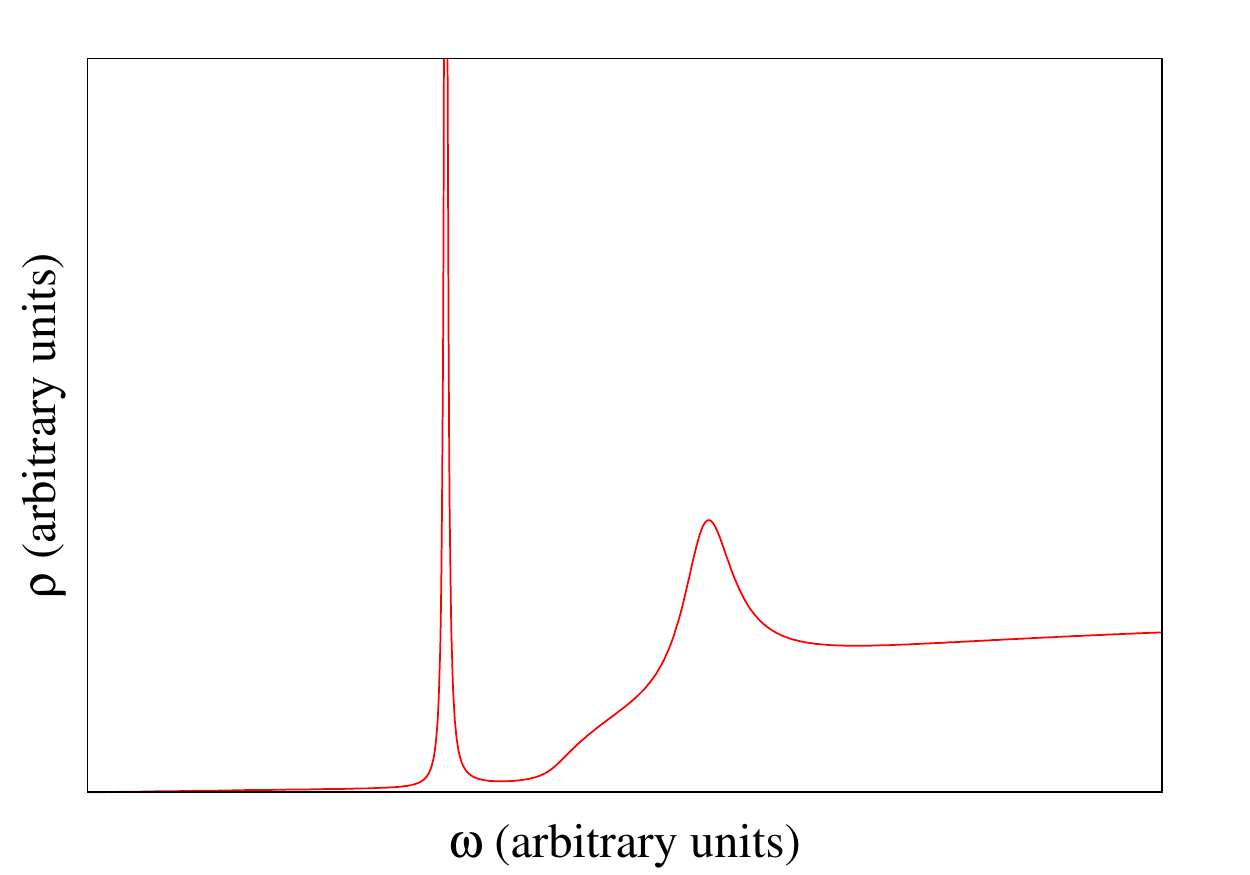}
  \caption{Schematic plot of a generic spectral function with two
    quasiparticle peaks for a given momentum.}
  \label{fig:schematic}
\end{figure}

Nevertheless, we have some generic expectations.  At zero temperature,
in a system where all the excitations are stable and massive, the
spectrum contains a stand-alone Dirac-delta peak, and, after a gap,
the continuum starts beyond a threshold value. In this case we
have an asymptotic state. At finite temperature or in the presence of
zero mass excitations there is no gap in the spectrum, but we
should count with the appearance of one or eventually more broadened
quasiparticle peaks and a multiparticle continuum (cf. Fig.~\ref{fig:schematic}).
The quasiparticle peak and the multiparticle continuum are generally not disjunct.

In order to characterize a ``typical'' spectrum we use trial spectra
resembling the above generic picture. We take the following
characteristic values: quasiparticle masses, quasiparticle
widths which depend on the height of the continuum at
the quasiparticle, the threshold value and relative heights of
quasiparticle peaks and the continuum. The absolute height is
determined by the sum rule, but it drops out from the expression of
pressure \eqref{eq:pressure}. All of these parameters can be momentum
and temperature dependent. Now we take these effects one-by-one,
compute the pressure coming from eq.~\eqref{eq:pressure} and compare
it to the pressure of a free gas. Quantitatively we will introduce the
effective number of degrees of freedom as
\begin{equation}
  \label{eq:Neff}
  N_{eff}(T) = \frac{P(T)}{P_0(T)},
\end{equation}
where $P(T)$ is the actual pressure from \eqref{eq:pressure} and $P_0$
is the free gas pressure.

%%%%%%%%%%%%%%%%%%%% MOM DEP %%%%%%%%%%%%%%%%

\subsection{Momentum dependence}
\label{sec:momdep}

First we start from the simplest approach to the realistic excitations
of the QCD plasma which is beyond the free particle HRG model. Here we
assume that the excitations are free particles, but they feel their
environment. Since at finite temperature the plasma singles out a rest
frame, a propagating hadron or parton will decay on this thermal
background depending on their spatial momenta. This effect results in
the spatial momentum dependence of the spectral functions
\eqref{eq:rho}. In particular in the crossover temperature regime, the
high and low momentum spectra can differ significantly.

Although we do not know exactly the spatial momentum dependence of the
spectral functions, we still bring physical arguments how it could
behave qualitatively. At large spatial momenta, corresponding to small
spatial distance, the thermal effects must be small, and we expect to
encounter quark- and gluon-like propagations. On the other hand at
small momenta, large scales, we should observe mostly hadronic
excitations.

To have a handle on this phenomenological expectation, we study here
an oversimplified model. In this model we assume a spectral function
which consists of a single Dirac-delta for low spatial momenta, and it
is zero\footnote{To be able to fulfill the sum rule, the spectral
  function should not be zero, but it can be so broad and shallow that
  it does not contribute to the pressure,
  cf. Section~\ref{sec:QPwidth}.}  for large spatial momenta. This is
a model for the hadronic modes. We use a sharp cutoff $\Lambda_c$
between the two regimes, but this will not be crucial for the results.

We can calculate the pressure from \eqref{eq:pressure}, but, since the
spectral function is so simple, we can perform the integrals, and
obtain the formula similar to the free gas pressure:
\begin{equation}
  \label{eq:momcut_g}
  \frac{P}{T^4} = -\frac1{2\pi^2} \int\limits_0^{g}\!
  dx\, x^2 \, \ln(1-e^{-\omega}),\qquad g=\beta \Lambda_c,\quad
  \omega^2=x^2+(\beta m)^2.
\end{equation}
For $m=0$ and $\Lambda_c\to\infty$ this provides the usual
Stefan-Boltzmann limit $\pi^2/90$. If $\Lambda_c\to0$, of course, the
pressure is zero.

\begin{figure}[htbp]
  \centering
  \includegraphics[height=5cm]{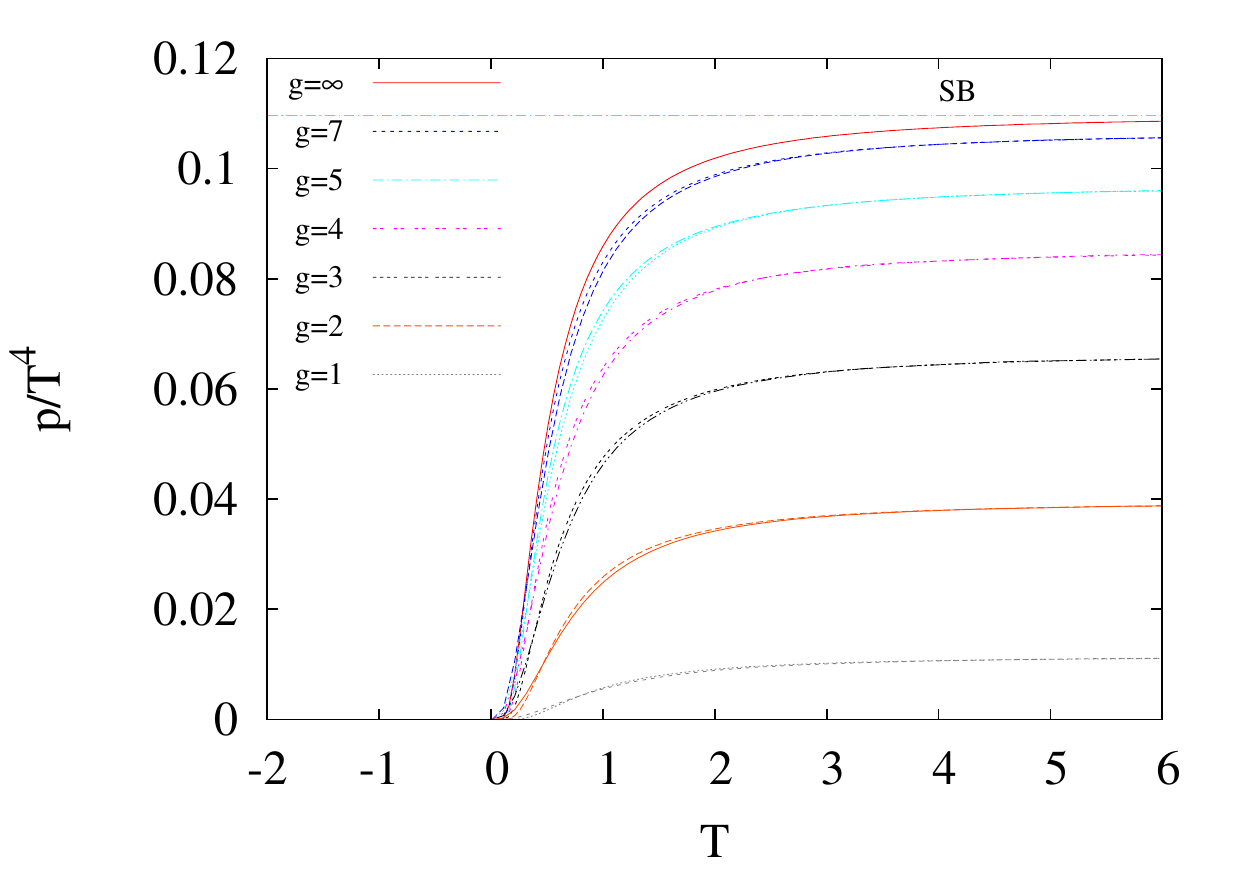}
  \caption{Pressure curves coming from applying an effective spatial
    UV cutoff in the pressure integral $\Lambda_c=gT$, and
    $N_{eff}P_0(m_{eff},T)$. These two curves cover each other up to a
    few percent. $SB$ indicates the Stefan-Boltzmann limit.}
  \label{fig:momcut_pressure}
\end{figure}

\begin{figure}[htbp]
  \centering
  \includegraphics[height=5cm]{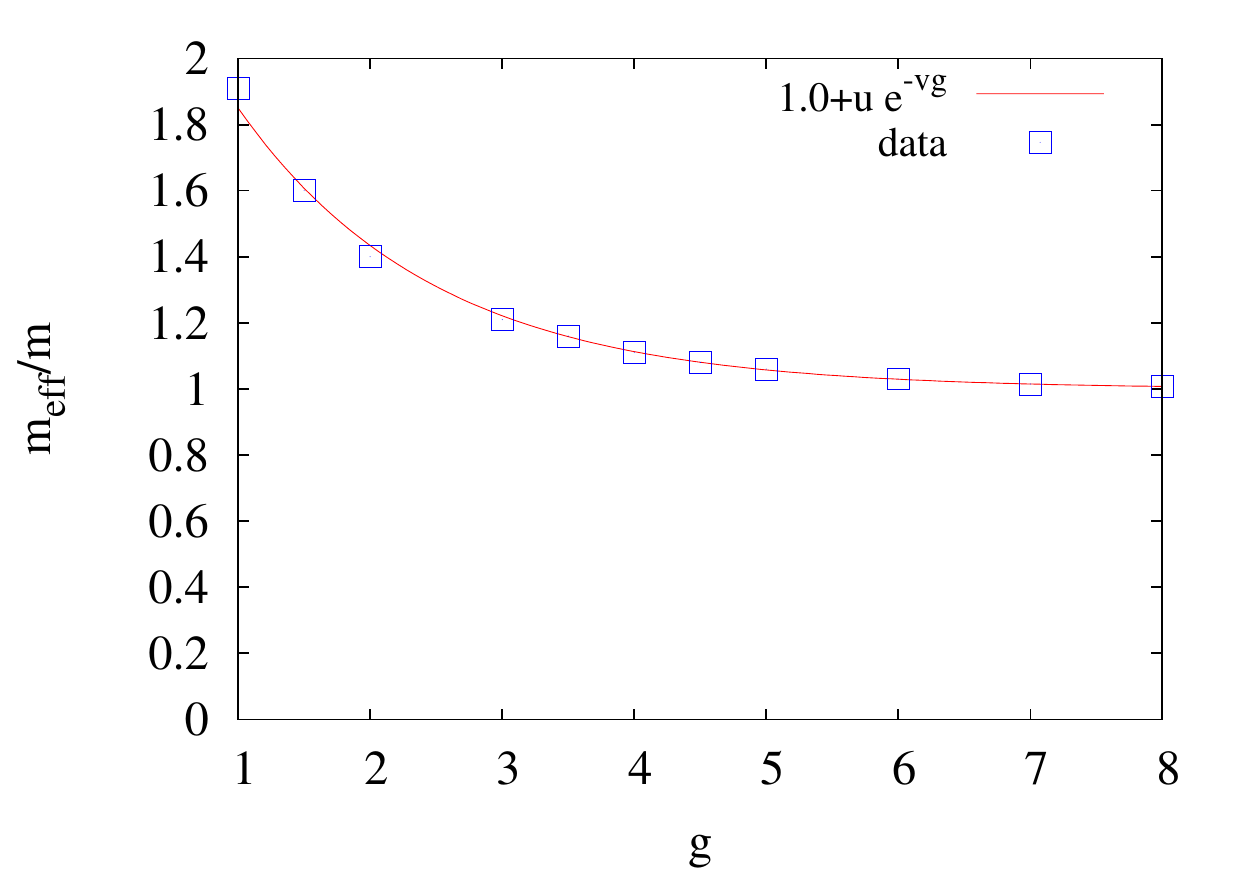}
  \includegraphics[height=5cm]{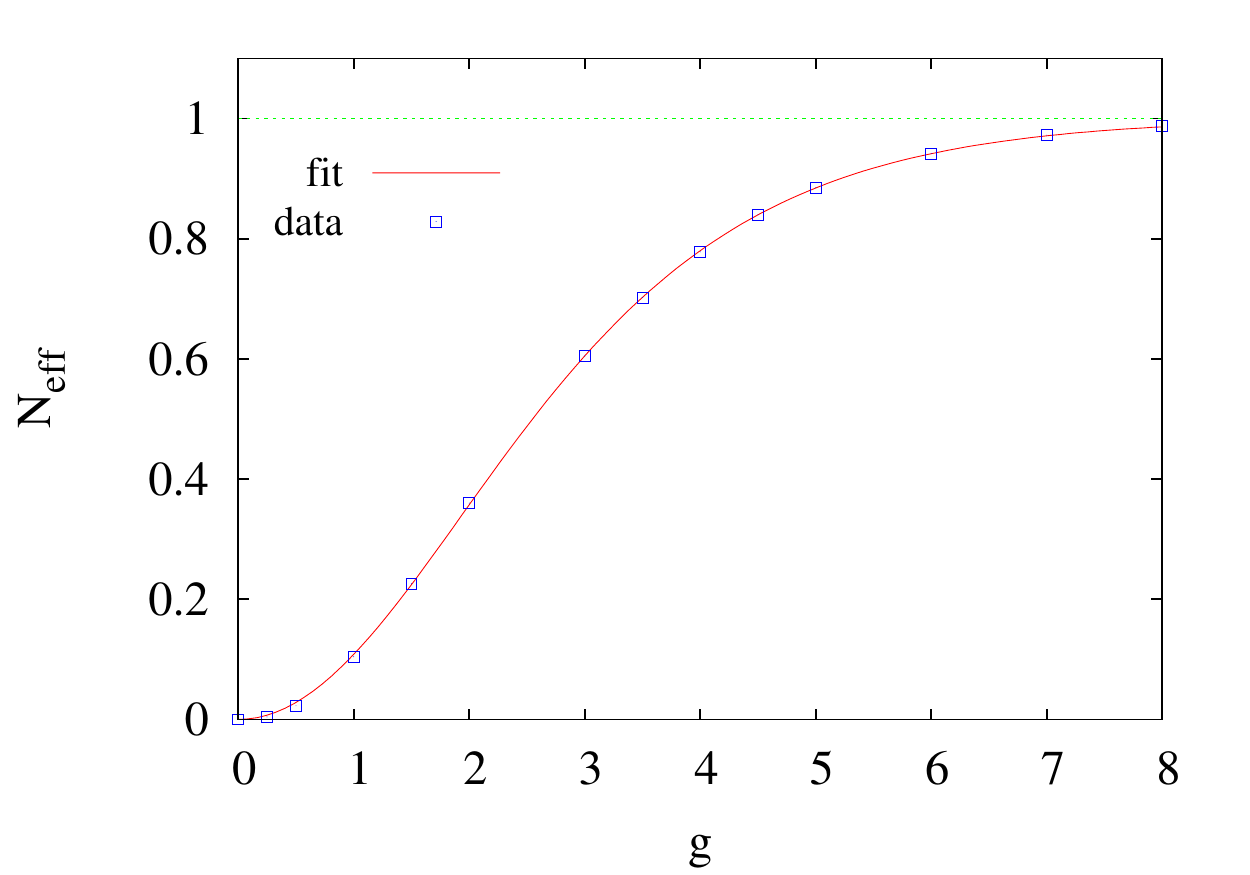}
  \caption{The effective mass parameter $m_{eff}$ (left figure) and
    the effective numbers of degrees of freedom $N_{eff}$ (right
    figure), together with some fit functions (cf. \eqref{eq:Nefffit}
    and \eqref{eq:fitneff}).}
  \label{fig:momcut_params}
\end{figure}

For the sake of simplicity, we analyze the above formula for constant
$g$ values. Since we did not use the temperature independence of $g$,
we are free to choose its $T$-dependence later. The advantage of the
constant $g$ choice is that the resulting $P(T)$ curves are very
similar to the ideal gas pressure curve with some effective mass
parameter and an overall suppression factor. In
Fig.~\ref{fig:momcut_pressure} these cutoff model curves are
presented, almost covering the fitted
\begin{equation}
  \label{eq:Nefffit}
  P(T) = N_{eff}(T)P_0(m_{eff}(T),T)
\end{equation}
curves.
One inspects an intriguing agreement: in fact even the largest
deviation between these two curves (which occurs when the suppression
is the largest at $g=1$), expressed by the relative difference
$(P-P_0)/P_{SB}$ does not exceed the few per cent level. We plot the
fit parameters in Fig.~\ref{fig:momcut_params}.

We can also try to give fitting functions to these plots, with the
sole aim to help to parametrize the numerically observed behavior of
the curves. For the effective mass we choose $m_{eff}/m = 1 + u
e^{-vg}$ where the best fits come from $u = 1.66,\,v =0.67$. The
function which fits best to the $N_{eff}$ data is given by
\begin{equation}
  \label{eq:fitneff}
  \mathrm{fit \: \:}  f(x) = \frac1{1+ x^{-2} e^{-(bx)^a}},\qquad x=\frac
  g{g_0},\qquad g_0=2.95,\quad a=1.79,\quad b=0.58.
\end{equation}
If the spatial momentum cutoff is temperature independent, then we
have to choose $g = \Lambda_c/T$. For large temperatures the exponent
can be neglected, and effectively we obtain $N_{eff}\sim (1+T^2)^{-1}$ 
like dependence.

%%%%%%%%%%%%%%%%%%%% MOMENTUM DEP 2 %%%%%%%%%%%%%%%%%%%%%%

\subsubsection{Momentum difference distribution}
\label{sec:momdifdistr}

So far we assumed that the elementary excitations propagate in the
matter which singles out a local rest frame, and so the spectral
function may depend on the momentum of the particle. But physically
the situation is more complicated. Taking a simple model where we have
binary collisions between free particles, the cross section in fact is
sensitive to the momentum difference between our particle and its
colliding partner. Statistically this means a momentum difference
distribution, which then leaves a trace in the spectral functions,
and, consequently in thermodynamics. To understand better the
statistical background physics, here we calculate this effect in the
limit of non-interacting massless particles \cite{Biro:2013rqa}.

\newcommand{\pd}[2]{{\frac{\partial #1}{\partial #2}}}

\newcommand{\be}{\begin{equation}}
\newcommand{\ee}[1]{\label{#1} \end{equation}}
\newcommand{\ba}{\begin{eqnarray}}
\newcommand{\ea}[1]{\label{#1} \end{eqnarray}}
\newcommand{\vs}{\vspace{5mm}}
\renewcommand{\c}[1]{{\cal{#1}}}

To begin with, we determine the distribution of {\em momentum differences} between
the elementary excitations. We assume free particles with one-particle
distribution function $f(E)$ where $E$ denotes their energy. Since the
particles are massless, $E=|\p|$. Using the relativistic kinematics for
pairs of massless particles we find
\be
 Q^2 =  -(p_1-p_2)^2 = 2E_1E_2(1-\cos\theta).
\ee{KINEM}
The distribution of $Q^2$ values is given by
\be
 P(Q^2) = \frac{\int dE_1 dE_2 d\cos(\theta) \, E_1^2E_2^2 f(E_1) f(E_2) \delta\left(Q^2-2E_1E_2(1-\cos\theta)\right)}{\int dE_1 dE_2 d\cos(\theta) \, E_1^2E_2^2 f(E_1) f(E_2)}.
\ee{THERMDIST}
Utilizing the Dirac delta functional for the integral over $\cos\theta$ this can be written as
\be
 P(Q^2) = \frac{\int_0^{\infty} dE_1 \int_{Q^2/4E_1}^{\infty}dE_2  \, \frac{1}{2}E_1E_2 f(E_1) f(E_2) }{\int_0^{\infty} dE_1 \int_0^{\infty} dE_2 \, 2E_1^2E_2^2 f(E_1) f(E_2)}.
\ee{PQ2}
This value is always between zero and one, its integral with respect
to $Q^2$ is one, due to the construction by
eq.(\ref{THERMDIST}). Since the thermal parton distribution, $f(E)$,
is a monotonic decreasing function, the numerator is maximal at
$Q^2=0$. 
This maximal value is given by
\be
 P(0) = \frac{1}{4} \frac{\int dE_1 \int dE_2 E_1E_2f_1f_2}{\int dE_1 dE_2 E_1^2E_2^2 f_1 f_2}
  =  \exv{\frac{1}{2E}}^2 = \frac{c^2}{T^2},
\ee{PQMAX}
with some constant $c$ depending on the distribution.

\begin{figure}[htbp]
  \centering
  \includegraphics[height=5cm]{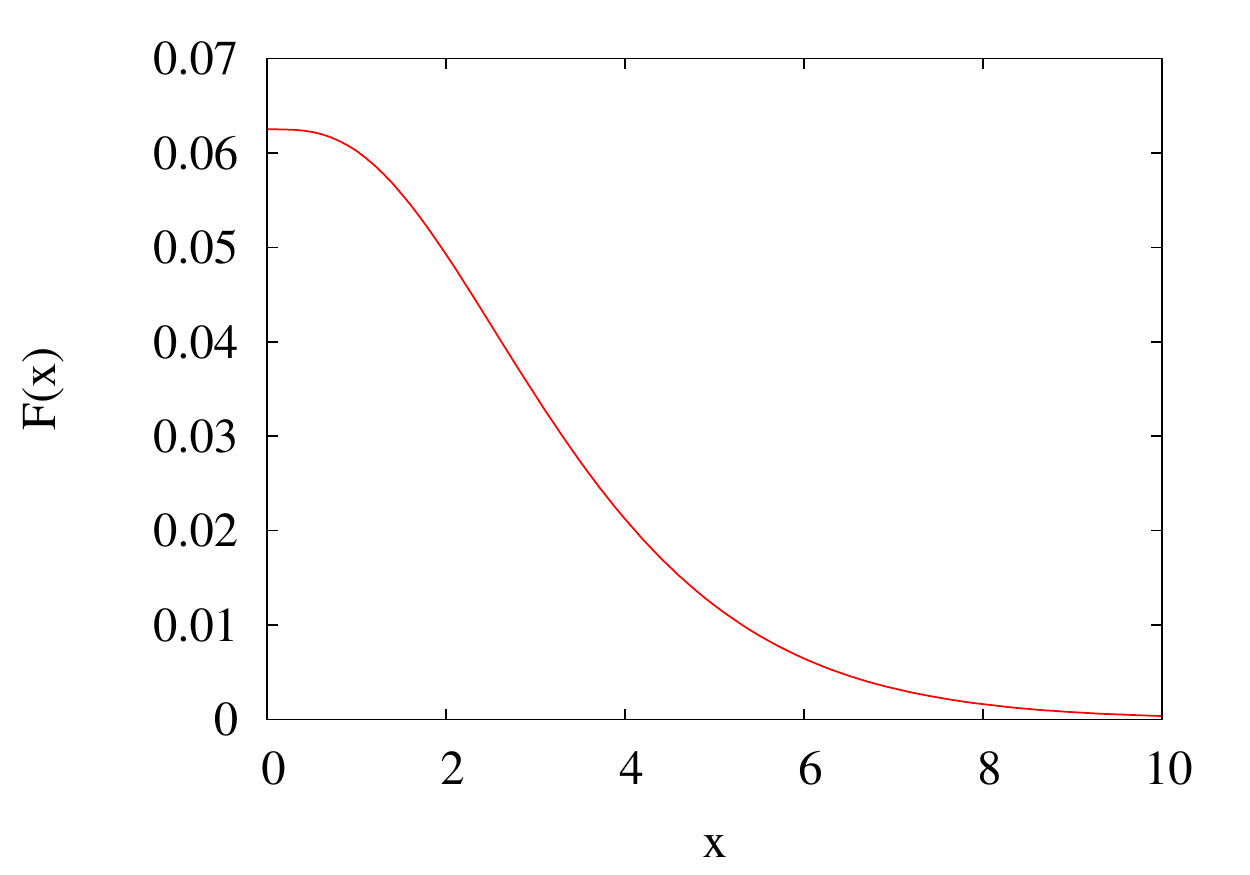}
  \includegraphics[height=5cm]{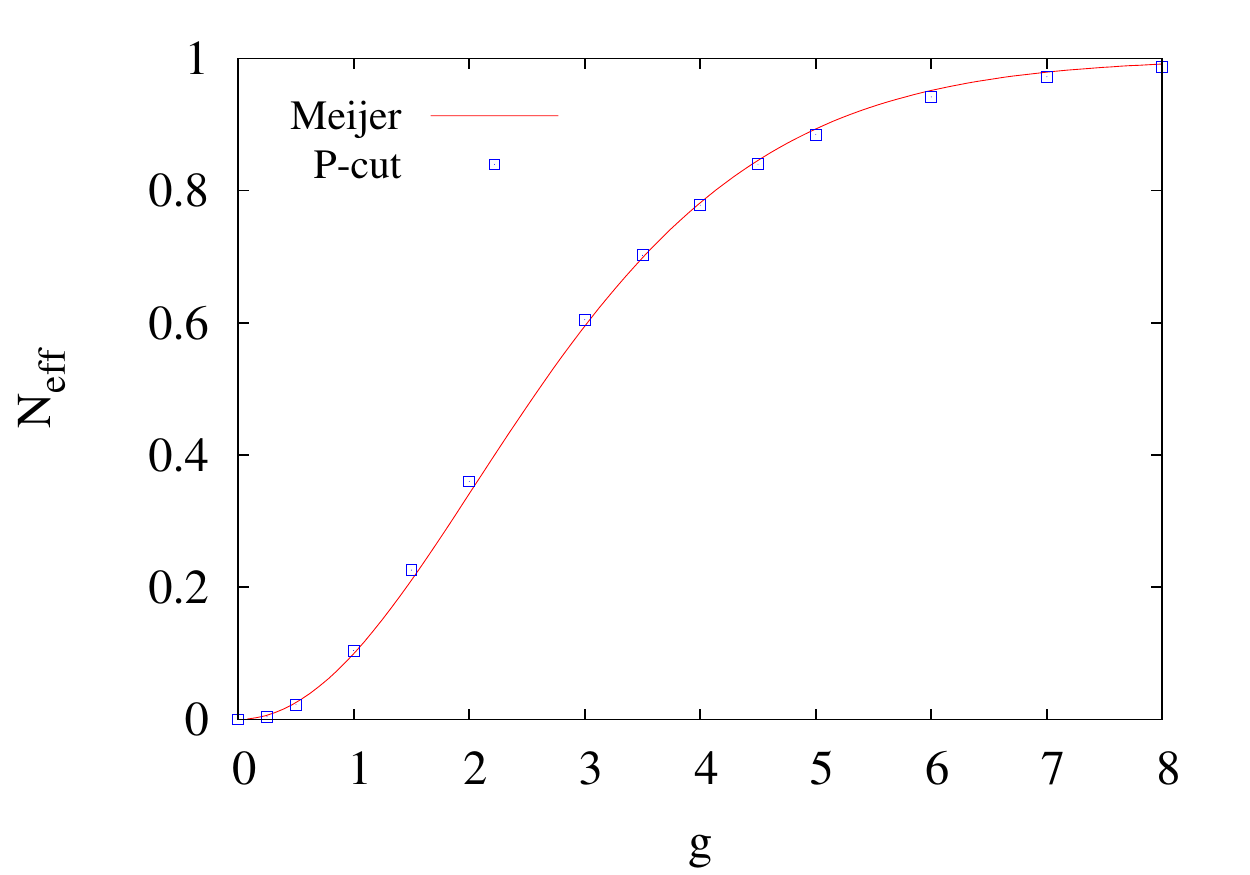}
  \caption{The left panel shows the momentum difference distribution
    $F(x)$ defined in \eqref{ORDER}. On the right panel the rescaled
    effective numbers of hadronic degrees of freedom
    (cf. eq.~\eqref{eq:Meijer}) is plotted as a function of $g$, together
    with the similar quantity defined in \eqref{eq:momcut_g}.}
  \label{fig:momcut_params_Meijer}
\end{figure}

For the Boltzmann distribution, $f(E)=e^{-E/T}/Z$, this distribution
can be obtained in analytic form:
\be
 P(Q^2) =  \frac{1}{T^2} F(Q^2/T^2),\qquad F(x) = \frac{1}{64} \left(
   x^{3/2} K_1(\sqrt{x})+2xK_2(\sqrt{x}) \right).
\ee{PQANAL}
The resulting rescaled momentum difference distribution is shown in the
left panel of Fig.~\ref{fig:momcut_params_Meijer}. Although this
result was derived for zero mass, we hope that for massive particles
this distribution is similar.

As it was already explained earlier in this section, the hadronic
modes have small width for small momentum transfer and large width for
large momentum transfer, while the partonic excitations behave in an
opposite way. Therefore the pressure contribution of the hadronic
modes with small momentum transfer is large, for large momentum
transfer is small. So the contribution for the effective number of
degrees of freedom $N_{eff}$ must also behave similarly. Just like in
the previous model of this section, we apply an extreme approximation
to this behavior, and we weight the momentum difference distribution
by a cut-off theta-function $\Theta(\Lambda^2-Q^2)$. Using our earlier
notation\footnote{Here $g$ can be temperature dependent.} $g=\beta\Lambda$,
the approximation for the hadronic $N_{eff}$ reads
\be 
N_{eff}^{hadr} = \int\limits_0^{g^2T^2} d{Q^2} \, P(Q^2)
=\int\limits_0^{g^2}\!dx\, F(x)
\ee{ORDER} 
by using the integration variable $x=Q^2/T^2$. This is the integrated
distribution function of the thermal $Q^2$ distribution. For small $g$
the distribution is nearly constant, so we can use \eqref{PQMAX} to
conclude that
\be
 N_{eff}^{hadr} \approx c^2 g^2.
\ee{HIGHT}
For Boltzmann distribution $c=1/4$.

The integral can be performed analytically, the formal result is
expressed with help of the Meijer G-functions:
\begin{equation}
  \label{eq:Meijer}
  N_{eff}^{hadr}(g) = \frac1{4} \left[ G^{2,1}_{1,3}\left(\frac
      g2,\frac12 \left| \begin{array}[c]{l} 1\cr 1,3,0\cr
          \end{array}\right.\right)
      +
      G^{2,1}_{1,3}\left(\frac
      g2,\frac12 \left| \begin{array}[c]{l} 1\cr 2,3,0\cr
          \end{array}\right.\right)
    \right] 
\end{equation}
This result is plotted in Fig.~\ref{fig:momcut_params_Meijer}. In
order to demonstrate the excellent agreement between the functional
forms we compare the rescaled $N_{eff}^{hadr}(0.78g)$ against
$N_{eff}$ coming from the previous analysis cutting the pressure
formula by excluding the large momentum part (cf. \eqref{eq:momcut_g}).
This surprisingly good agreement between the results of two vastly
different approaches needs explanation. 
Although in both models the pressure calculation involved a momentum
cut-off, soft part for hadrons, hard part for partons, 
the fundamental assumptions differ: the single-particle momentum
cut relative to the medium, while the $Q^2$-cut relative between
pairs of particles. Also Lorentz-transformation properties differ
essentially. Inspecting further Fig.~\ref{fig:momcut_params_Meijer} one
realizes that in the transition period, from the 90\% hadron dominated
phase to the 90\% parton dominated one, $g$ increases by a factor
of $6$. This nicely accords with the ratio $6$ between a parton
picture based $1$ GeV characteristic soft scale and the QCD lattice
transition temperature scale of $167$ MeV. Based on these two
coincidences we believe that the simple cut-off model performs
surprisingly well.

%%%%%%%%%%%%%%%%%%%% THERMAL MASS %%%%%%%%%%%%%%%%%%%%%%

\subsection{Thermal mass}
\label{sec:thermmass}

To further improve the model of the hadronic/partonic excitations, we
may take into account the effect of the medium on the particle
masses. Since MC simulations do not show strong temperature variation,
there is no signal that near $T_c$ the mass parameters would
diverge or behave any extraordinary way: the temperature variation of
the masses must not be modified near $T_c$.

\begin{figure}[htbp]
  \centering
  \includegraphics[height=5cm]{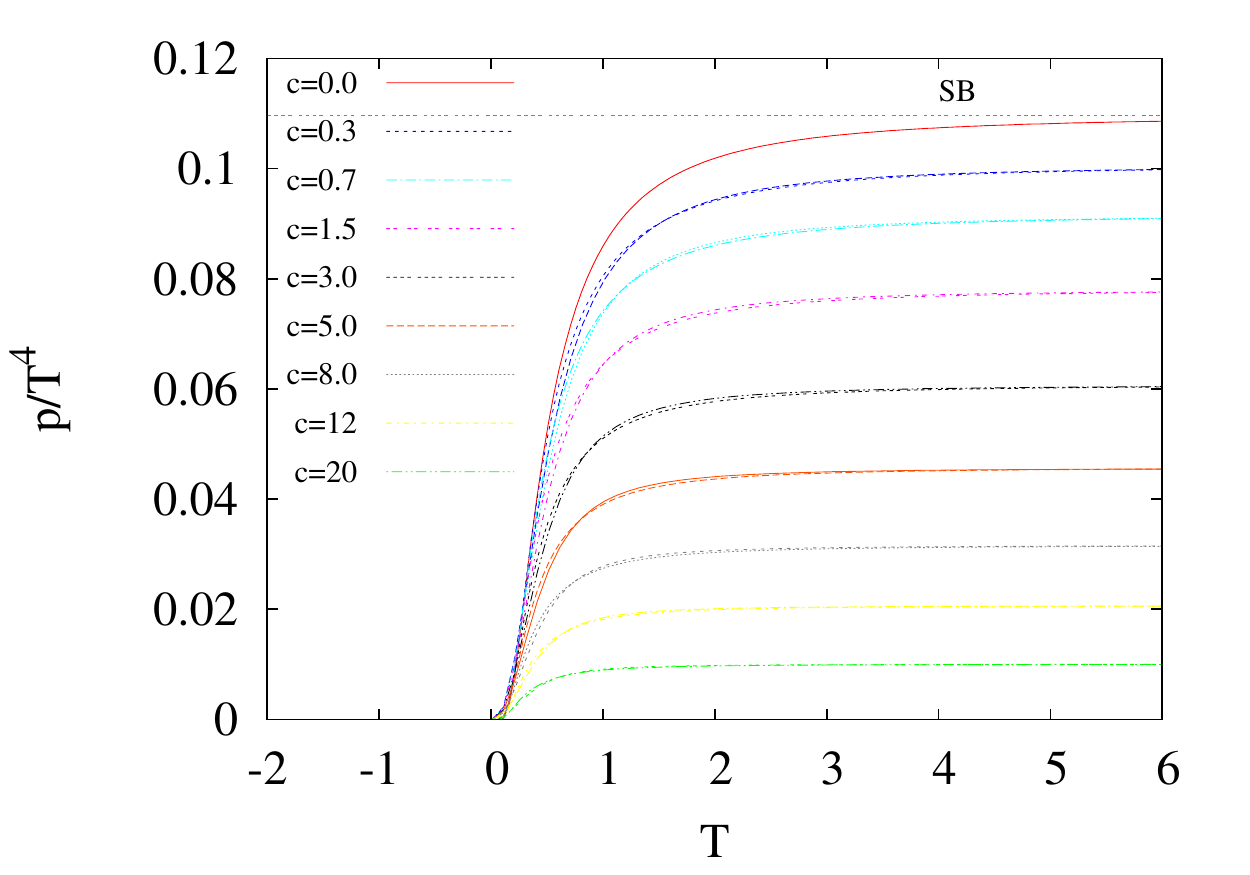}
  \caption{Pressures coming from using thermal mass $m^2=m_0^2+cT^2$,
    choosing $m_0=1$ units, and the $N_{eff}P_0(m_{eff},T)$
    assumption, respectively. The two curves almost cover each
    other. $SB$ indicates the Stefan-Boltzmann limit.}
  \label{fig:thermmass_pressure}
\end{figure}

\begin{figure}[htbp]
  \centering
  \includegraphics[height=5cm]{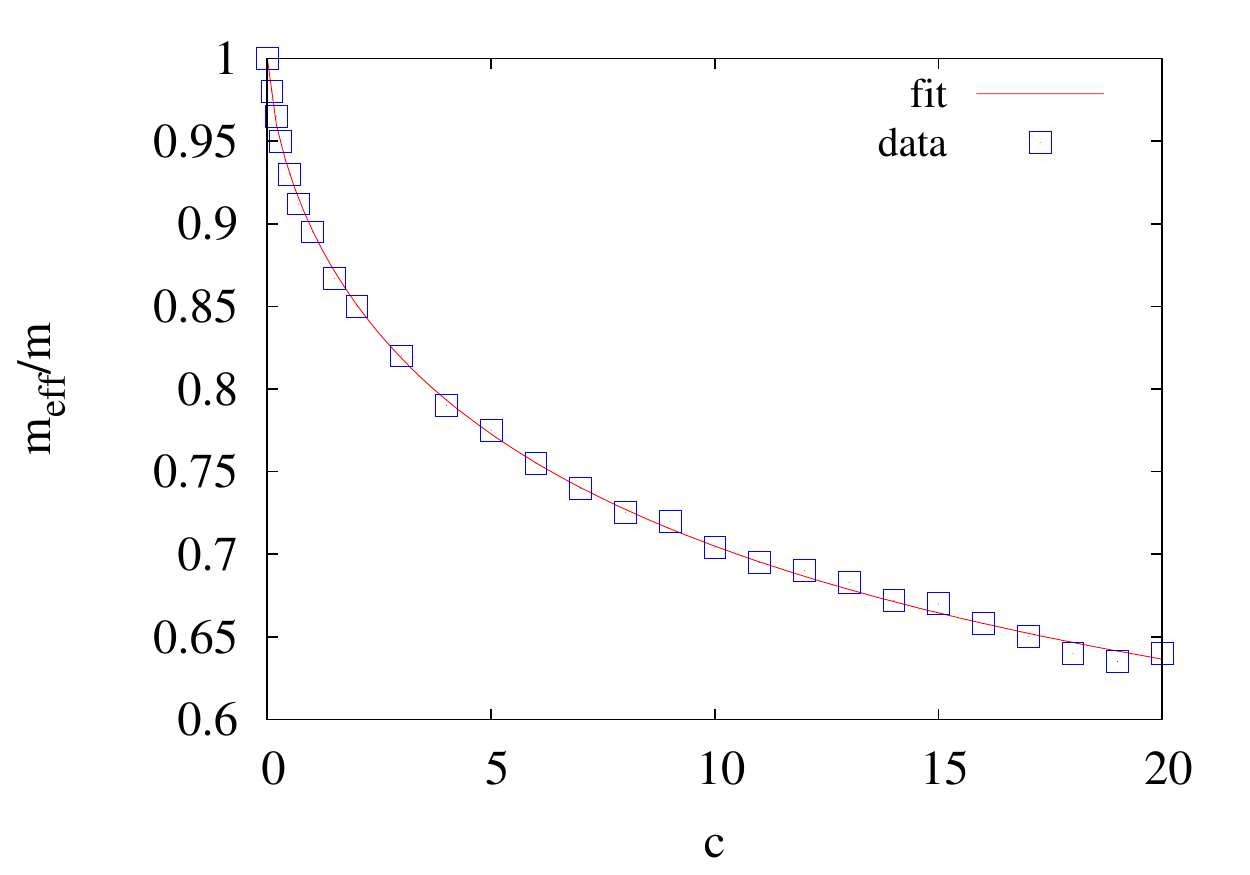}
  \includegraphics[height=5cm]{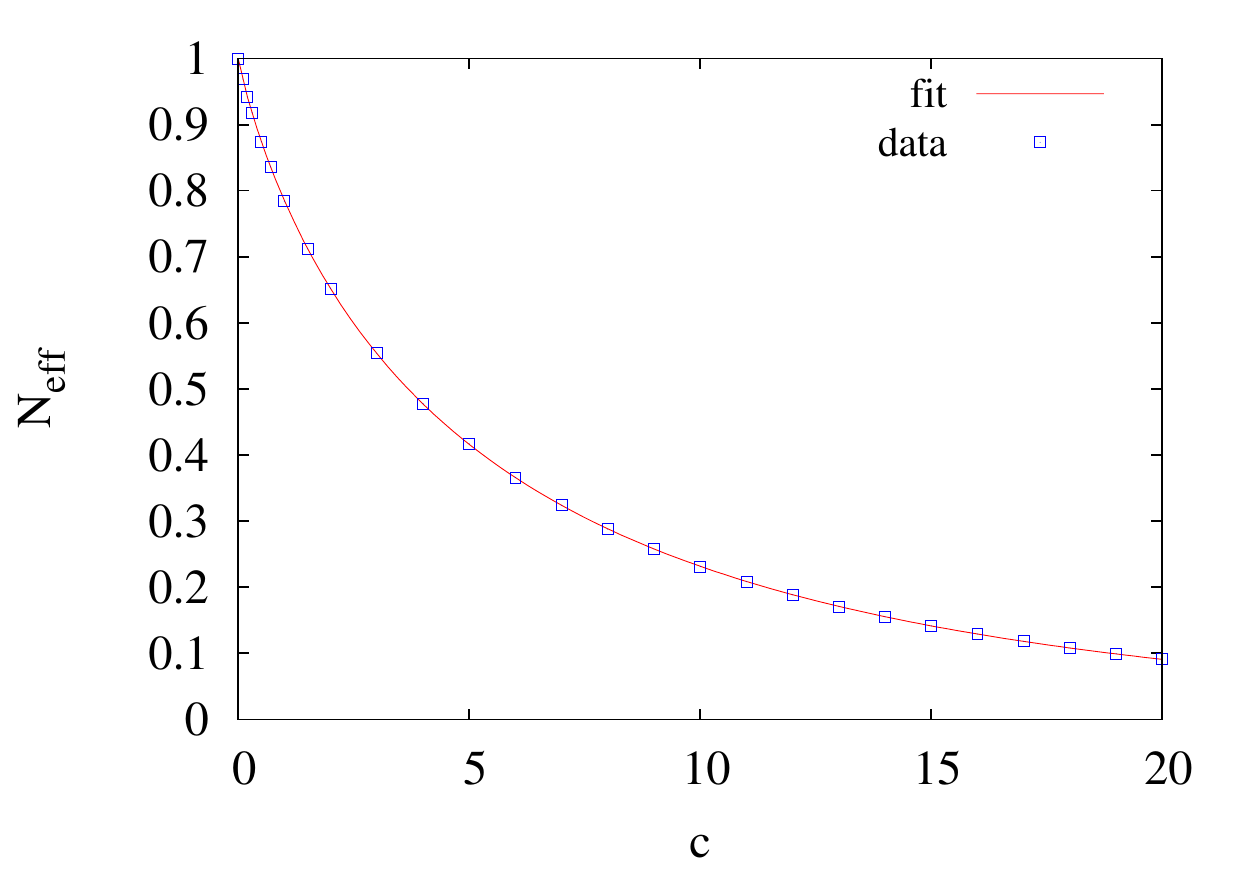}
  \caption{The fitted effective mass parameter (left figure) and the
    effective number of degrees of freedom (right figure), together
    with some fit functions (cf. \eqref{eq:fitmass1} and \eqref{eq:fitneff1} ).}
  \label{fig:thermmass_params}
\end{figure}

To describe the effect of the variation of the masses with the
temperature, we follow a similar path as in the previous
subsections. We realize that for a choice $m^2 = m_0^2 + c T^2$ the
pressure curve runs similarly to the free gas pressure, only an
effective number of excitations and an effective $m_0$ should be
introduced. This is demonstrated in Fig.~\ref{fig:thermmass_pressure}.
As before, we plot the pressure with thermal mass and the rescaled
free gas pressure (cf. \eqref{eq:Nefffit}) with some fitted effective
mass. The agreement is quite good also here, we find that the scaled
deviation, $(P-P_0)/P_{SB}$, is at most on the percent level. We also
display the fit parameters in Fig.~\ref{fig:thermmass_params}.
The fit function for the effective mass reads as
\begin{equation}
  \label{eq:fitmass1}
  \frac{m_{eff}}{m}\biggr|_{fit} = 1-\frac{wx^v}{1+ux^v},\qquad u =
  0.208,\quad v = 0.666,\quad w = 0.125,
\end{equation}
while the fit for $N_{eff}$ was performed with
\begin{equation}
  \label{eq:fitneff1}
  N_{eff}^{(fit)} =  \frac{e^{-ax}}{1+bx^c},\qquad a = 0.0527,\quad b =
  0.208,\quad c = 0.871.
\end{equation}

What we can observe is that for a constant $c$ value the pressure
curves do not reach the Stefan-Boltzmann limit. We expect that $c$
decreases at most logarithmically, therefore in the temperature range
up to $1-2$ GeV the constant $c$ approximation must be appropriate.

%%%%%%%%%%%%%%%%%%%%% QUASIPARTICLE  %%%%%%%%%%%%%%%%%%%%%%%%%%%%%%%%%%%%

\subsection{Quasiparticle width and continuum height}
\label{sec:QPwidth}

We take an approximation for the spectral function neglecting the
momentum dependence and discuss the pressure in terms of
the quasiparticle width and wave function renormalization. 
Because of a fixed sum rule, the quasiparticle wave
function renormalization is related to the continuum height.

\begin{figure}[htbp]
  \centering
  \includegraphics[height=5cm]{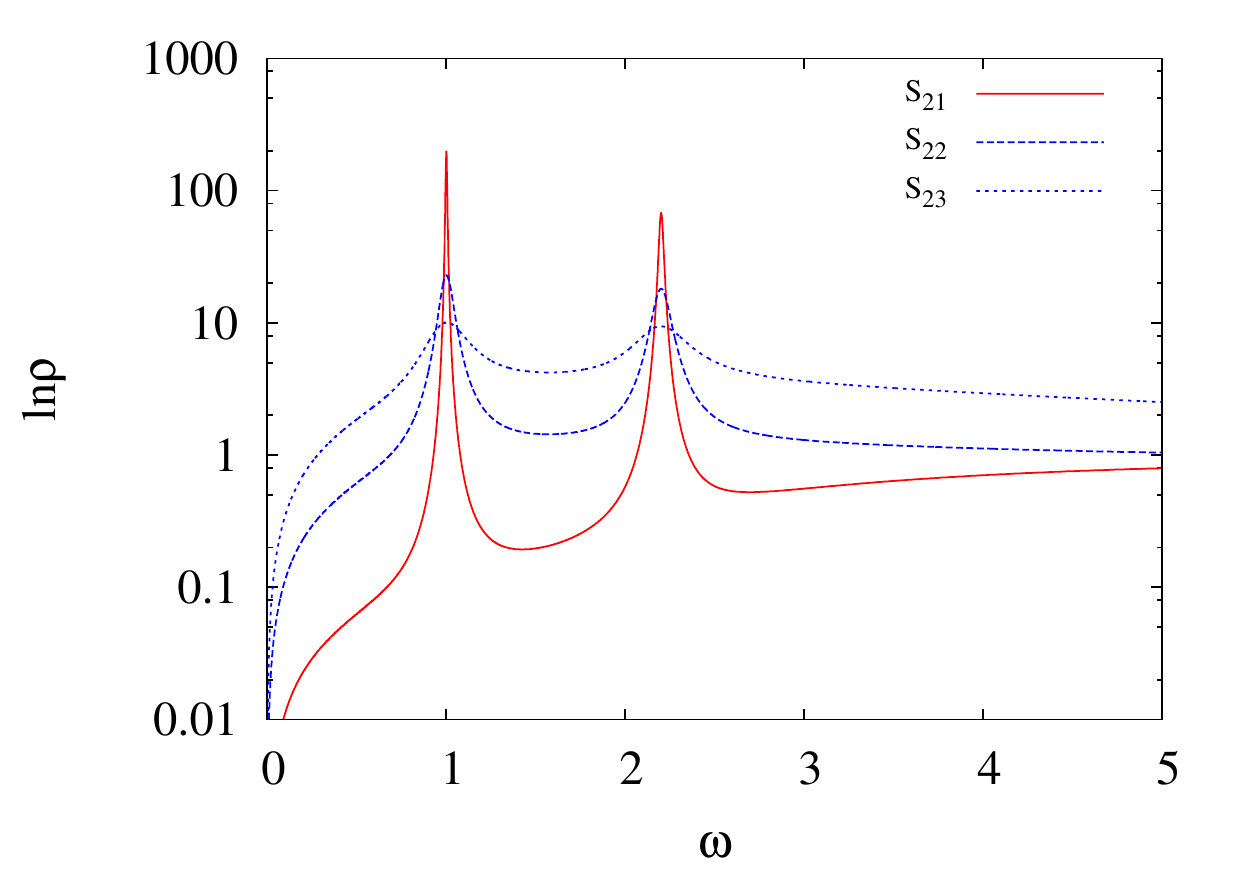}
  \includegraphics[height=5cm]{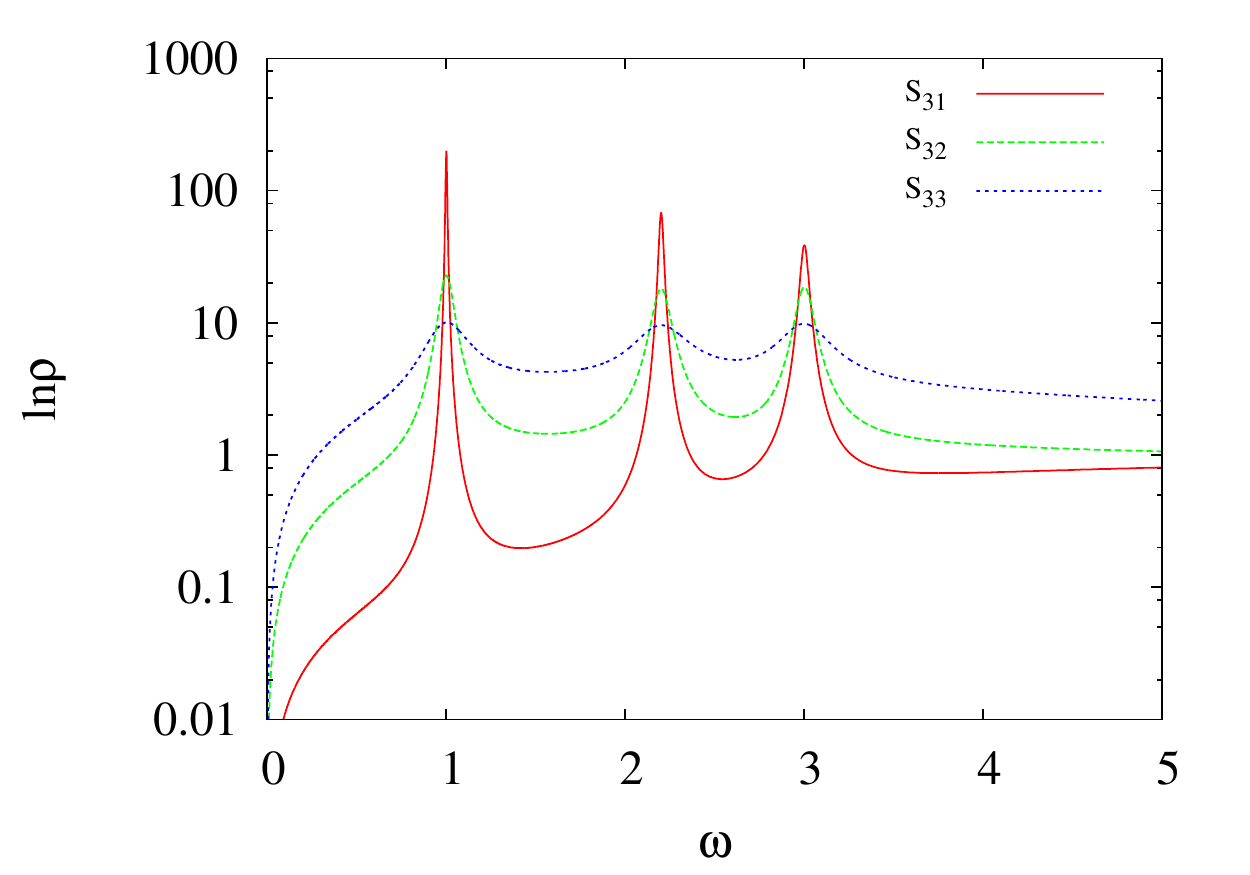}
  \caption{Spectra with two and three quasiparticle peaks and a
    continuum. The width parameters of the plots can be found in
    Table~\ref{tab:params}.}
  \label{fig:samplespectra}
\end{figure}

This situation is partly discussed in Ref.~\cite{Jakovac:2013iua}. 
Now we examine more realistic spectra: these contain several (two or three) quasiparticle
excitations above a multiparticle background. The multiparticle
contribution is modeled by a two particle cut with imaginary threshold
value (for more motivation see~\cite{Jakovac:2002rc,Jakovac:2013iua})
\begin{equation}
  \rh_{cont}(p) = \frac p{p^2+m_{th}^2} \Im\sqrt{m_{th}^2+iS-p^2}.
\end{equation}
This function is nowhere zero, it mimics the effect of finite
temperature spectral functions or spectral functions with several
threshold values. We assume that one quasiparticle has a mass below
$m_{th}$, so it would be a stable excitation if $S=0$ were true; the remaining ones
have masses above $m_{th}$. The widths of the
quasiparticle peaks are determined by the background, and all peaks
have the same area. We tested several spectra under such conditions,
some sample functions are shown in Fig.~\ref{fig:samplespectra}.
The parameters for the samples can be found in Tables~\ref{tab:params}
and \ref{tab:params1}.
\begin{table}[htbp]
  \centering
  \begin{tabular}[c]{||c||c|c|c|c|c|c|c|c||}
    \hline\hline
    name & $S_{21}$ &$S_{22}$ & $S_{23}$ & $S_{24}$ &$S_{25}$ & $S_{26}$ &$S_{27}$ & $S_{28}$ \\
    \hline
    $\gamma_1$ & 0.005 & 0.046 & 0.138 & 0.010 & 0.022 & 0.032 & 0.046
    & 0.066\\
    $\gamma_2$ & 0.015 & 0.059 & 0.172 & 0.020 & 0.032 & 0.043 & 0.059
    & 0.083 \\
    \hline\hline
  \end{tabular}
  \caption{The parameters of the plots with two peaks.}
  \label{tab:params}
\end{table}
\begin{table}[htbp]
  \centering
  \begin{tabular}[c]{||c||c|c|c|c|c|c|c|c||}
    \hline\hline
    name & $S_{31}$ &$S_{32}$ & $S_{33}$ & $S_{34}$ &$S_{35}$ &
    $S_{36}$ &$S_{37}$ & $S_{38}$ \\
    \hline
    $\gamma_1$ & 0.005 & 0.046 & 0.138 & 0.010 & 0.022 & 0.038 & 0.055
    & 0.079\\
    $\gamma_2$ & 0.015 & 0.059 & 0.172 & 0.020 & 0.032 & 0.050 & 0.070
    & 0.100 \\
    $\gamma_2$ & 0.026 & 0.058 & 0.161 & 0.028 & 0.035 & 0.050 & 0.068
    & 0.095 \\
    \hline\hline
  \end{tabular}
  \caption{The parameters of the plots with three peaks.}
  \label{tab:params1}
\end{table}
The quasiparticle widths are determined by the continuum height at the
quasiparticle mass. These spectra are rather typical for hadronic
channels.

\begin{figure}[htbp]
  \centering
  \includegraphics[height=5cm]{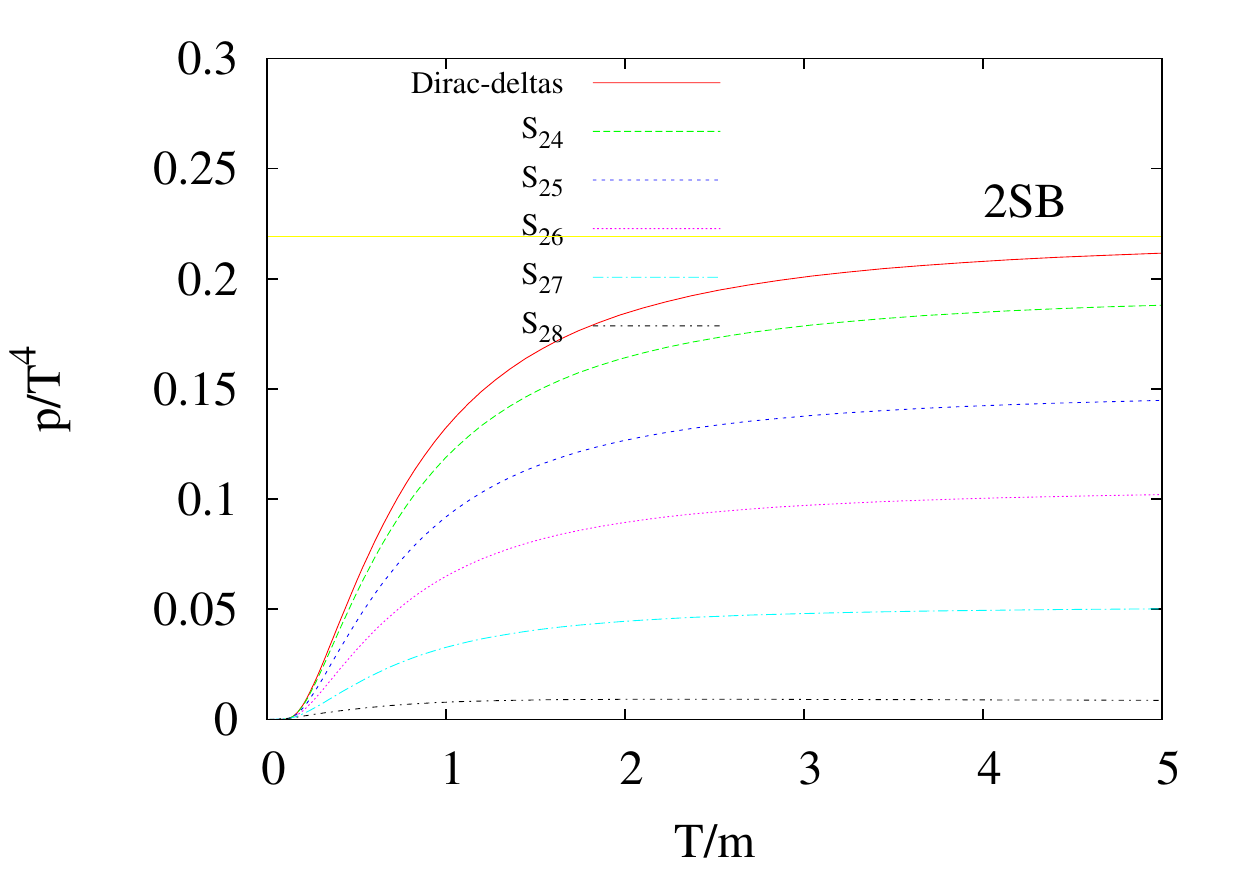}
  \includegraphics[height=5cm]{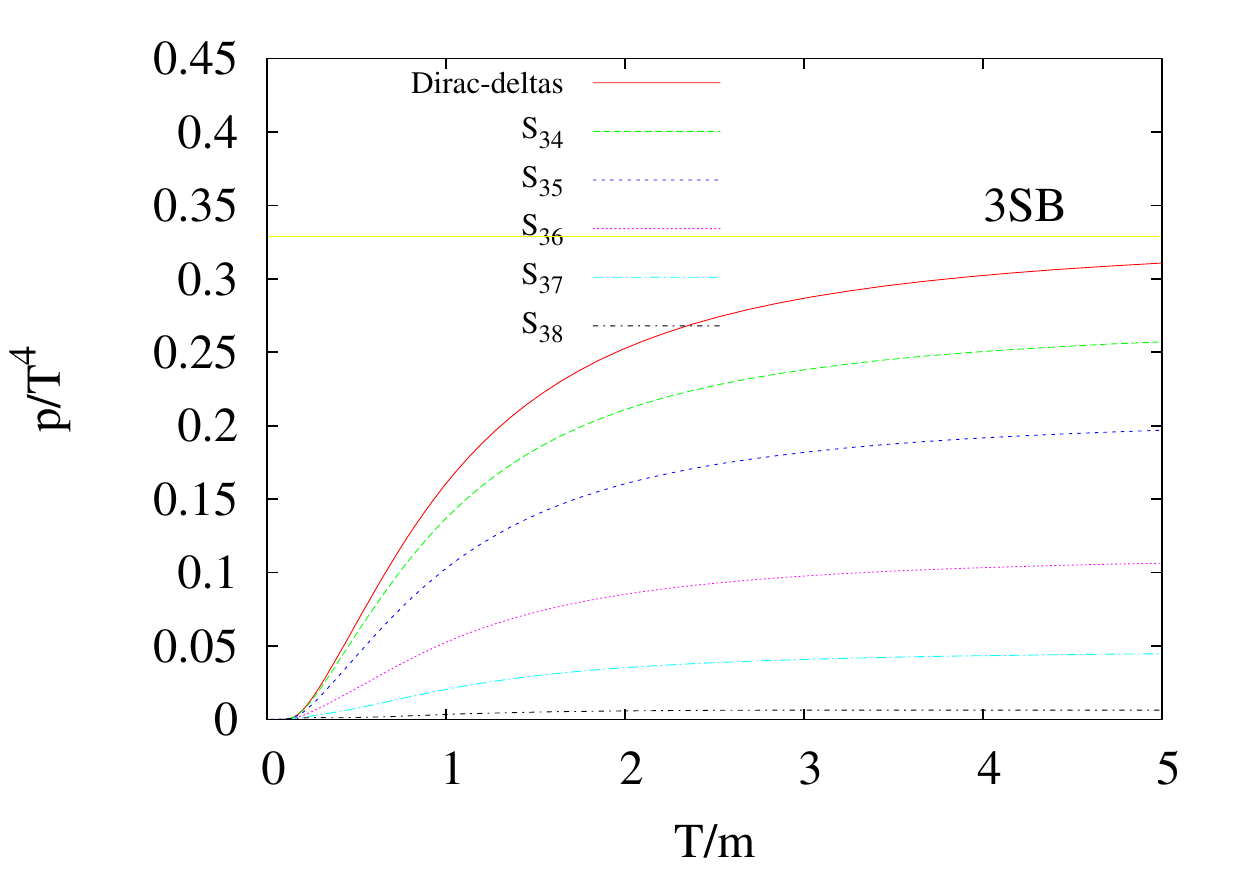}
  \caption{Pressure coming from the general spectra with two peaks
    (left figure) or three peaks (right figure). $SB$ denotes
    the Stefan-Boltzmann limit.}
  \label{fig:pressures}
\end{figure}

One inspects the pressure, based on eq.~\eqref{eq:pressure}, 
belonging to these spectra on Fig.~\ref{fig:pressures}

\begin{figure}[htbp]
  \centering
  \includegraphics[height=5cm]{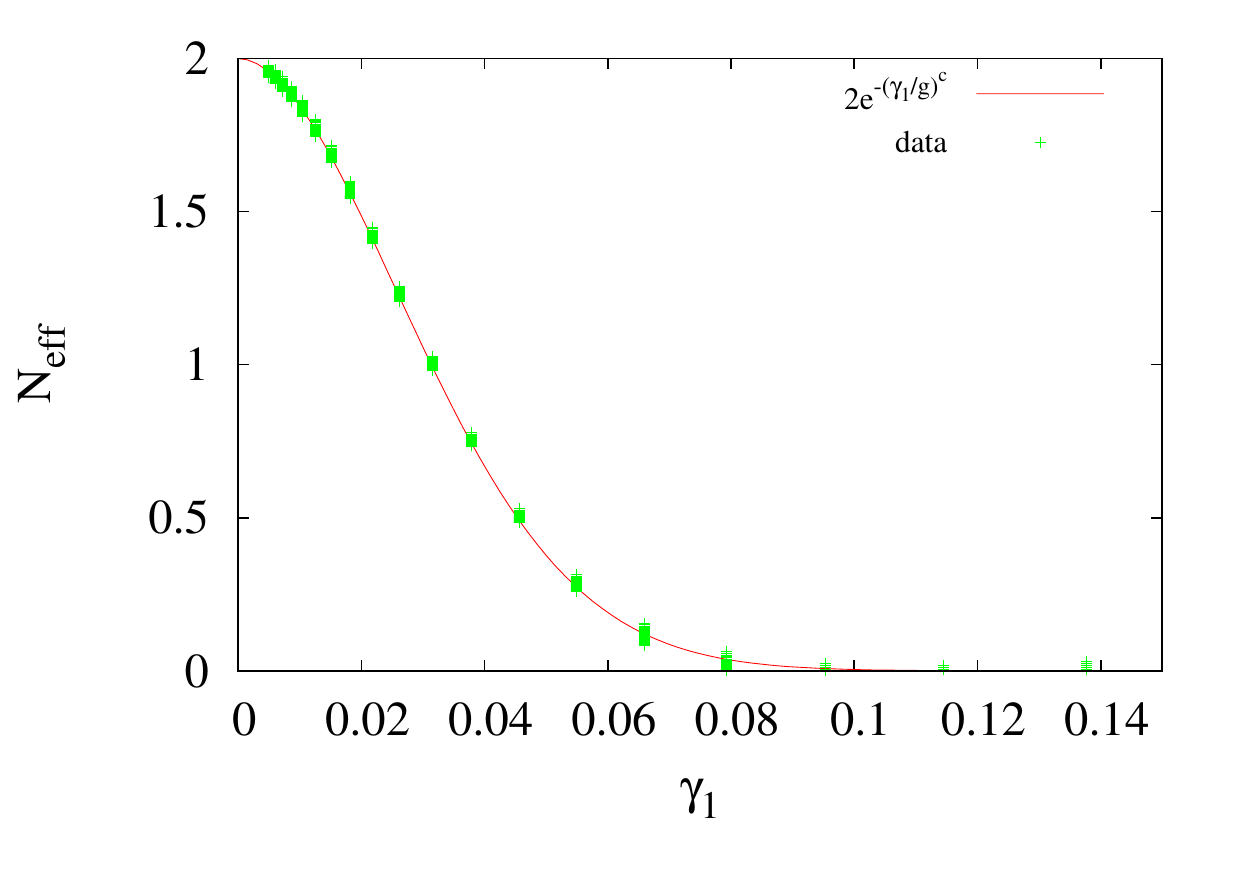}
  \includegraphics[height=5cm]{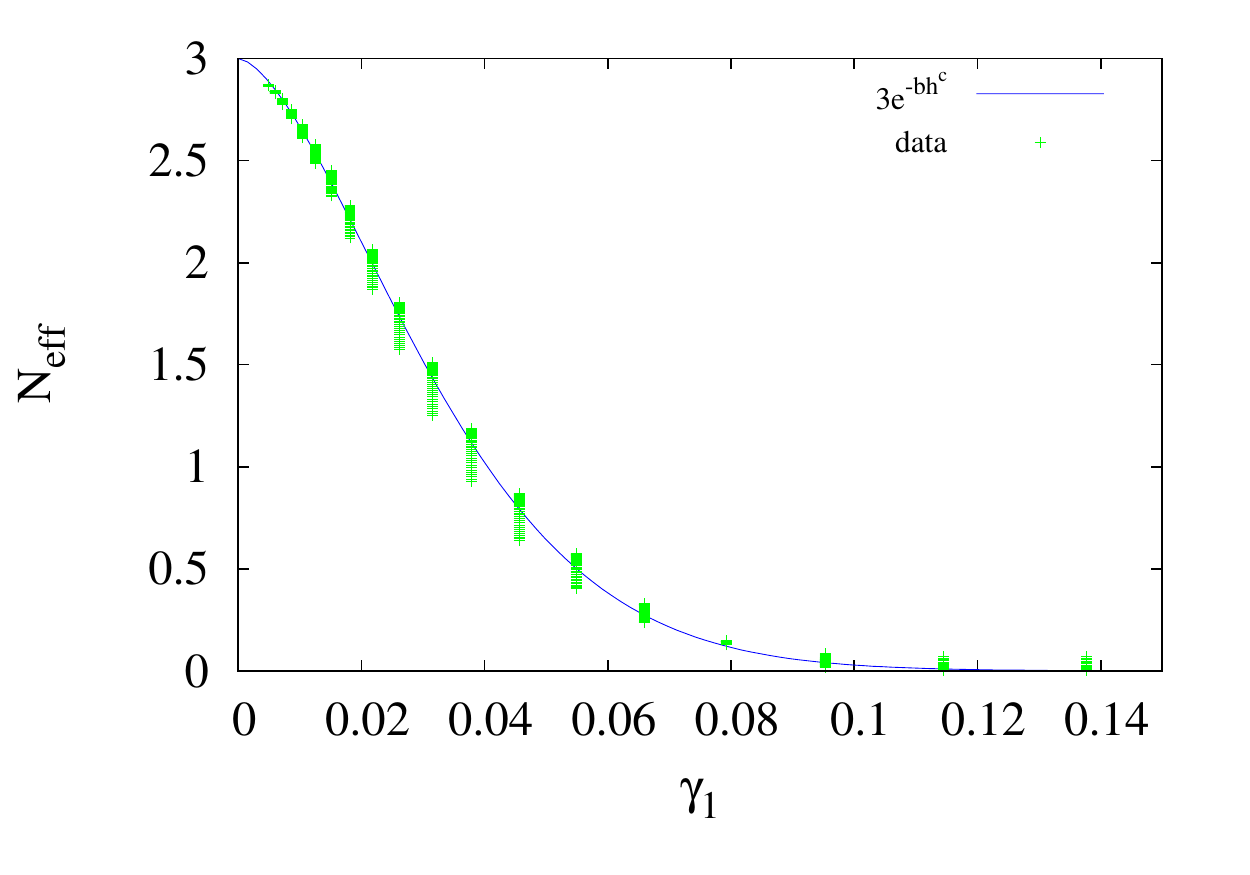}
  \caption{The effective number of dof for spectra with two peaks
    (left figure) or three peaks (right figure). In the horizontal
    axis we see the width of the first peak, which is
    proportional to the height of the continuum. Data taken from
    \cite{Borsanyi:2010cj}.}
  \label{fig:neff}
\end{figure}

Finally we can read out the effective number of dof defined in
\eqref{eq:Neff}, see Fig.~\ref{fig:neff}.
The widths of the data points correspond to the temperature variation
of $N_{eff}$. We see that, in agreement with the experience with the single
quasiparticle case in~\cite{Jakovac:2013iua}, $N_{eff}$ is almost
temperature independent. A single curve can be fitted through all
points. For the single quasiparticle case a Gaussian was a good choice
(cf.~\cite{Jakovac:2013iua}), here this Ansatz does not perform nicely.
Instead we consider a stretched exponential fit function
\begin{equation}
  N_{eff} = \exp(-(\gamma_1/g)^c),
\end{equation}
where $\gamma_1$ is the width of the first peak, proportional
to the height of the continuum.  This Ansatz produced fits with
excellent agreement with the data. The fit parameters can be seen in
Table~\ref{tab:fitp}.
\begin{table}[htbp]
  \centering
  \begin{tabular}[c]{||c||c|c||}
    \hline
    \# of peaks & g & c \\
    \hline
    2 & 0.038 & 1.88 \\
    3 & 0.038 & 1.60 \\
    \hline
  \end{tabular}
  \caption{Parameters of the $N_{eff}$ fit $\exp(-(\gamma_1/g)^c)$.}
  \label{tab:fitp}
\end{table}
As it can be seen, the less peaks the closer is the exponent, $c$, to 
the Gaussian ($c=2$) reached for a single peak.

%%%%%%%%%%%%%%%%%% QCD THERMO %%%%%%%%%%%%%%%%%%%%%%

\section{QCD thermodynamics}
\label{sec:QCDthermo}

In the previous section we studied several typical spectral
functions, and computed the pressure stemming from them. As we have
seen, with an appropriate parametrization of the modification effects,
we could factor out an effective number of degrees of freedom
$N_{eff}$. It is, in a fair approximation, temperature
independent. For free particles it is $N_{eff}=1$, in the interacting
case it is $N_{eff}\le1$. It is a complicated object, partly because
the different effects discussed in the previous sections result in
complicated fit functions, partly because in reality the fit
parameters are temperature dependent themselves in a nontrivial
way. On the other hand, to approach QCD thermodynamics, the hadronic
degrees of freedom should be treated statistically, so only the robust
properties count.

One of the most prominent property of $N_{eff}$ is that it can be a
constant for large temperatures. Finite width, presence of the
continuum, reduction of the spatial phase space and thermal mass
effects all may result in this phenomenon.

The next robust property is that $\ln N_{eff}$ may depend on the
parameters as a power law. The dependence of $N_{eff}$ on the
quasiparticle width (or continuum height), on the phase space
exclusion and on the thermal mass parameter all contain
exponential factors. In a realistic case the width
usually grows with the temperature. The momentum cutoff parameter is
either constant, or -- if the physical cutoff $\Lambda_c$ is constant
-- decreases with $T$. The thermal mass coefficient
($m^2=m_0^2+cT^2$) is more or less constant. That means that $\ln N_{eff}$
is dominated by the continuum background i.e. by width effects.

Talking about the temperature dependence of the continuum background
height we must mention two aspects. One is a generic thermodynamical
effect: the radiative corrections in general grow with the
temperature, thus the background height, together with the
quasiparticle width grows with the temperature. Guided by dimensional reasoning,
we shall assume the simple $\gamma^2 = \gamma_0^2+ \kappa T^2$
dependence. The zero temperature parton width $\gamma_0$ depends
on the hadron mass. We consider a simplified description: we
assume that there is a fast rise in $\gamma_0(m)$ meaning an
effective cutoff in the number of available hadrons~\cite{Cleymans:2011fx}. 
So in the calculations we just choose
$\gamma_0=0$ and the number of hadronic excitations goes to some $N$
in the range of $[3000-5000]$. Once $\gamma$ is given, we use $\ln N_{eff}
\sim \gamma^c$.

For the width of the partons (QGP degrees of freedom), we must take
into account another effect: the partonic cross section increases with the
presence of hadronic excitations. So we assume that the parton
width is proportional to some power of $N_{eff}^{hdr}$, the number
of hadronic excitations. It is crucial for the fittability of the MC pressure. 

To make life easier, we assume that the width of all hadrons have
approximately the same temperature dependence, and partons have another single
temperature dependent width, respectively. Moreover, we 
neglect the variation of the mass parameter of the free gas pressure
and the explicit temperature dependence of the partonic width. Then
the simplified Ansatz for the QCD pressure is given as the following:
\begin{eqnarray}
  P_{hadr}(T) &&= N_{eff}^{(hadr)}\hspace*{-1em}
  \sum\limits_{n\in\mathrm{hadrons}}^N\hspace*{-1em}
  P_0(T,m_n),\qquad \ln N_{eff}^{(hadr)} = (T/T_0)^b,\nn
  P_{QGP}(T) &&= N_{eff}^{(part)}\hspace*{-1em}
  \sum_{n\in\mathrm{partons}}\hspace*{-1em}
  P_0(T,m_n), \qquad \ln N_{eff}^{(part)} = G_0 + c (N_{eff}^{(hadr)})^d.
\end{eqnarray}
Here $P_0(T,m)$ is the pressure of a free gas with mass $m$ at
temperature $T$. For the masses of the hadronic excitations we assume
a Hagedorn-like spectrum~\cite{Hagedorn:1965st,Broniowski:2004yh}
\begin{equation}
  m_n = m_\pi + T_H\log n.
\end{equation}
The lowest mass $m_\pi$ should be in the order of the pion mass, the
parameter $T_H$ is the Hagedorn temperature. For the partons we assume
some effective masses, namely $m_{ud}=300$ MeV, $m_s=450$ MeV and in the
best fits we found $m_{gluons}=500-550$ MeV.

We have quite a few parameters at hand, but the pressure curve found in MC
simulations have some well defined regimes. It helps to find the
correct values for the parameters. First of all at very low
temperatures the pressure comes exclusively from the light hadrons,
which helps to find the fit value of $m_\pi$. We note here that all
hadrons are taken into account with unit multiplicity, therefore
$m_\pi$ is some ``smeared'' pion mass. Below $T\sim 160$ MeV the
pressure is still dominated by the hadrons, this helps to fit
$T_H$. The HRG pressure coming from the hadronic pressure with no
suppression would overshoot the real pressure near $\sim 160-200$
MeV. To avoid this, we adjust the fit parameters $T_0$ and
$b$. We learn that $b$ must be $1.5-2$ (cf. Table~\ref{tab:fitp}),
restricting further the allowed fit ranges.

At very large temperatures the pressure is determined by the QGP
degrees of freedom. Although there exist perturbative calculations
based on the QCD Lagrangian, but to be consistent, here we give an
effective, phenomenological description also to this regime. At large
temperatures the pressure is more or less constant. In principle there
is a slight, logarithmic increase in the pressure, but the lattice
data until 1.5 GeV are consistent with a constant (for larger
temperatures see \cite{Borsanyi:2012ve}). This will determine
$G_0$. As the temperature starts to decrease hadrons appear, from this
the coefficients $c$ and $d$ can be determined. Thus in the QGP sector
$G_0$ is determined by the MC data, there remain only two parameters
to fit.

\begin{figure}[htbp]
  \centering
  \includegraphics[height=5cm]{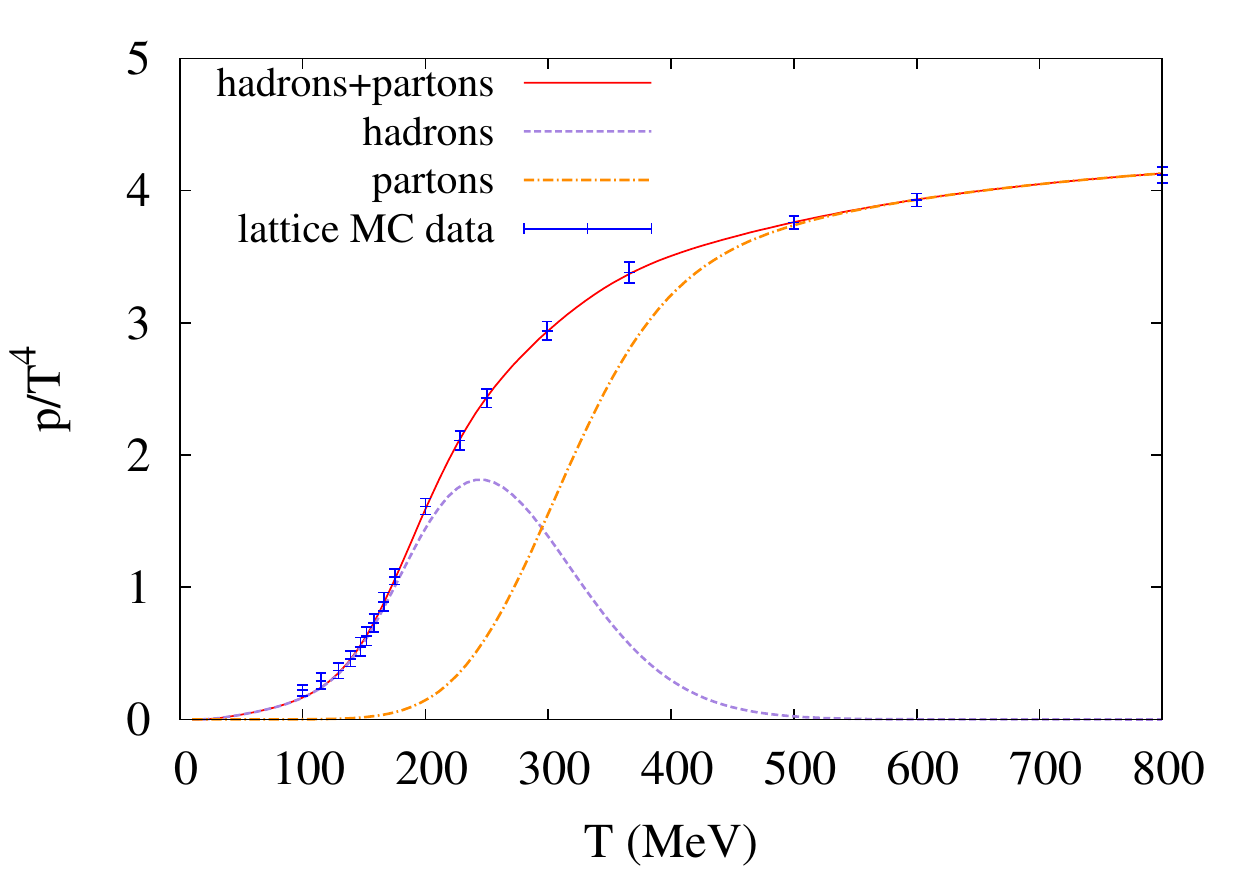}
  \includegraphics[height=5cm]{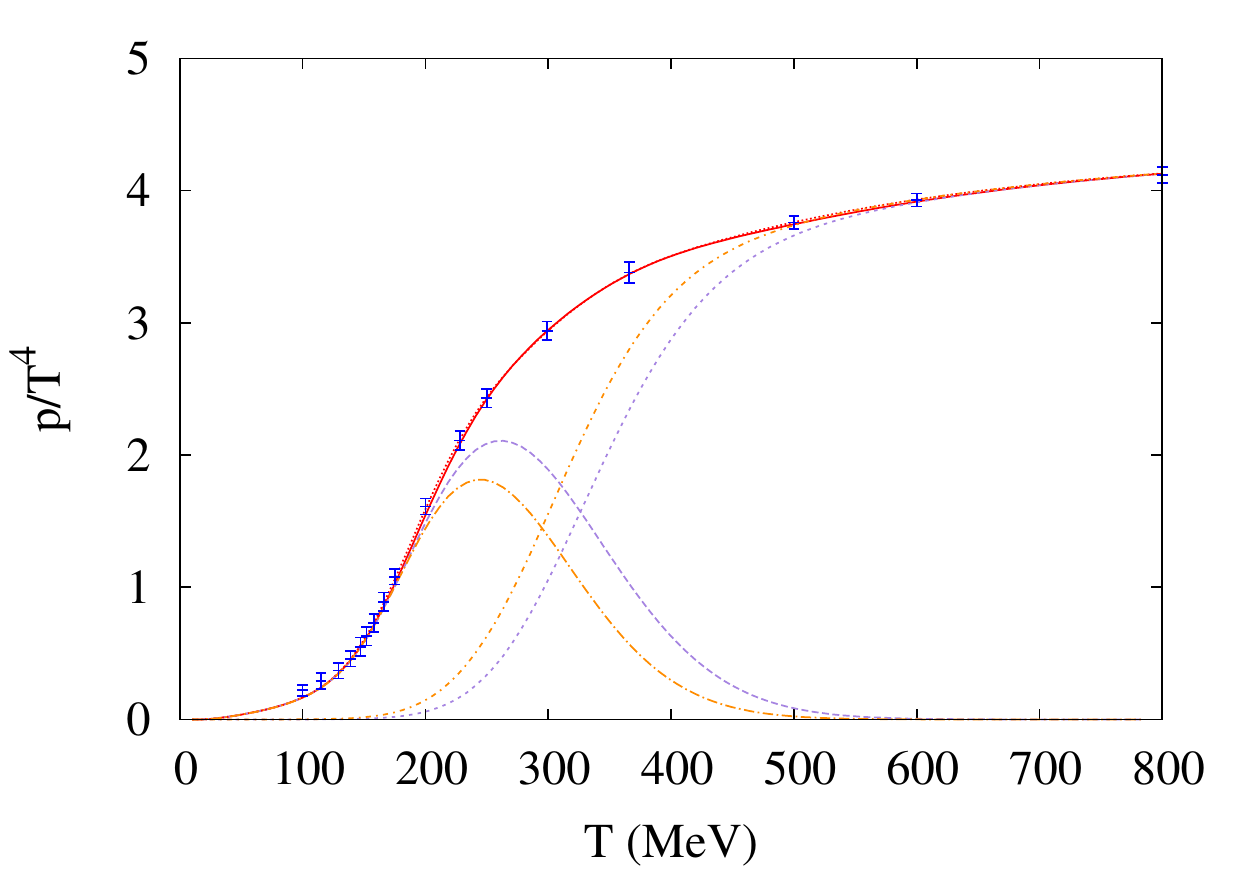}
  \caption{Pressure of the QCD fitted by our model. The fit curve goes
    through all points. On the left panel a single selected fit is
    shown, on the right panel we compare results of two different
    fits.}
  \label{fig:QCDpressure}
\end{figure}

The best fit for the pressure goes through all the points, as it is
shown by the left panel of Fig.~\ref{fig:QCDpressure}.
We also made further fits, two fits are shown on the right panel of
Fig.~\ref{fig:QCDpressure}. The fit parameters are summarized in Table~\ref{tab:QCDfits}
\begin{table}[htbp]
  \centering
  \begin{tabular}[c]{||c|c|c|c|c|c|c|c|c|c||}
    \hline
    nmr. & $N_H$ & $m_\pi$ & $T_0$ & $T_H$ & $d$ & $b$ & $c_g$ & $c_q$ & $G_0$ \\
    \hline
    1 & 5000 & 140 & 165 & 187 & 0.9 & 1.95 & 22 & 9 & 0.157 \\
    2 & 3000 & 120 & 190 & 195 & 0.9 & 2 & 9.7 & 9.7 & 0.151 \\
    \hline
  \end{tabular}
  \caption{Fit parameters for fitting lattice MC data.}
  \label{tab:QCDfits}
\end{table}

%%%%%%%%%%%%%%% SUB: QCD PRESSURE %%%%%%%%%%%%%%

\subsection{Properties of the QCD pressure}
\label{sec:discussion}

As the curves on Fig.~\ref{fig:QCDpressure} show, with the simple
hadron+parton model above we could very accurately fit the lattice MC
results for the QCD pressure. Despite the fact that we have some fit
parameters, this is not self-evident. Most of the parameters of
Table~\ref{tab:QCDfits} are not entirely free fit parameters, they are fixed
by some physical requirements. Still
there remain 5 or 6 parameters (depending on the fitting strategy)
which could be played with in order to obtain the best fit.

It was crucial for the fits, as we mentioned before, that the
fermionic suppression factor depends on the number of available
hadronic modes ($N_{eff}^{(hadr)}$). Without it one can not achieve
that partons with mass as low as 300-600 MeV, i.e. lower than most
hadronic excitations, would give significant contribution to the
pressure only at large temperatures.

There are some robust properties which was shared in all
fits we could do. First of all it was known before that up to
$T\approx T_c$ the hadron resonance gas alone can describe the full
QCD pressure. This we used as an input for the fits. But this also
means that the partonic pressure just starts to appear at this
temperature. Being a crossover transition the partonic pressure must
appear continuously. The question remains, however, that how long does
it take before the quarks dominate the thermodynamics.

The conventional view of the hadron-QGP transition considers it as being rather
fast, the ``width'' of the susceptibility curves give a hint for
that. Unfortunately this is not a strong argument, since for a single
free massive scalar field the $p/T^4$ curve exhibits very similar
properties than the QCD pressure curve, but there any kind of phase
transition is missing. 

According to the present calculation the rule of thumb for the QCD
pressure is that it is dominated by hadrons until $T\approx 2T_c$. The
hadronic and QGP contributions are equal approximately at $2T_c$,
later the quark degrees of freedom will dominate. Only at $T\approx
3T_c$ do we obtain a system that can be described solely by the QGP
excitations.

This finding is more or less is in accordance with the perturbative
studies~\cite{Haque:2014rua}. In these studies, according to the
authors, the pressure is described down to about approximately $2T_c$, where,
in the light of the present calculation, the GQP modes dominate the pressure.

%%%%%%%%%%%%%%%%%%%%%%%

\section{Conclusions}
\label{sec:conc}

In the present paper we proposed a model of the strong interactions
which is capable to give an account for the numerically determined QCD
thermodynamics. The basis of this model is a quadratic field theory
with a general spectral function. Once the model is fixed one can
calculate the pressure exactly.

For phenomenological applications we have to parametrize the spectral
function with some characteristic numbers such as the position (mass)
of the peak or peaks, the height and threshold behavior of the
continuum, the spatial momentum dependence for finite temperature or
density applications, and the sum rule. One can then determine the
pressure of the system as a function of these numbers. The most robust
finding of these studies is that the pressure of the system decreases
and eventually vanishes as the quasiparticle properties become less
and less enhanced. We have characterized this tendency by the
effective number of degrees of freedom $N_{eff}$, which is just the
ratio of the pressure coming from the actual spectral function and the
one with vanishing continuum. By an appropriate choice of the
parametrization one can achieve that $N_{eff}$ is approximately
temperature independent (cf. Figs.~\ref{fig:momcut_params},
\ref{fig:thermmass_params}, \ref{fig:neff}): then for each spectral
function is modeled by a single number. $N_{eff}$ is a function of the
parameter set which determines the spectral function. We have
determined this function for several plausible spectral functions and
proposed fitting formulas for it.

As a last step we applied our approach for a statistical mixed hadron and
parton model of QCD. Ingredients of this model are hadrons,
described by a statistical mass distribution and by common, average hadron
properties for the spectral function. The other ingredient is a parton
gas consisting of $u,d$ and $s$ quarks and gluons. While the hadronic
spectral functions are determined on their own, the quark spectral
functions, in particular the continuum height, must have depended on
the number of available hadrons $N_{eff}^{(hadr)}$. The parameters of
the model were adjusted to fit the lattice MC measurement of the QCD
pressure.

As a result we do not only fit the lattice MC pressure curve excellently, but we
also describe the hadron\,--\,parton decomposition of the QCD plasma at a
given temperature. We have found that for $T\lsim T_c\approx 156$ MeV
only hadrons are present in the medium. These hadrons do not
disappear at the would-be critical temperature: for $T_c\lsim T\lsim
2T_c$ they still dominate the pressure, although not in the original
hadron resonance gas form, but as excitations with broad spectral
functions. In the higher temperature regime the hadrons melt gradually,
for $2T_c\lsim T\lsim 3T_c$ they still give considerable
contribution to the total pressure. Only at $T\approx 3T_c$ do we
arrive at the point where the QCD pressure is dominantly given by the
fundamental QCD degrees of freedom.

As future prospects we plan to apply our description of QCD EoS to
determine further statistical observables of the strongly
interacting plasma like the transport coefficients. Moreover we consider to
apply this description also at finite chemical potential.

\section*{Acknowledgments}

The authors thank instructive discussions with A. Patk\'os. They also
acknowledges discussions with Zs. Sz\'ep, P. Mati, U. Reinosa. This
work is supported by the Hungarian Research Fund (OTKA) under contract
No. K104292 and K104260.


\begin{thebibliography}{99}
\bibitem{transportcoeff} transport coefficients are small in the weak
  interaction limit
\end{thebibliography}
\end{document}